\pdfoutput=1
\documentclass[referee]{gji}
\usepackage{timet,color}
\usepackage[utf8]{inputenc}
\usepackage[urlcolor=blue,citecolor=black,linkcolor=black]{hyperref}
\usepackage[german, english]{babel}
\let\bibhang\relax

\usepackage{natbib}
\usepackage{doi}

\usepackage{cancel}
\usepackage{multirow}
\usepackage{supertabular}
\usepackage{algorithmic}
\usepackage{algorithm}

\usepackage{amsthm}
\usepackage{amsmath}
\usepackage{float}
\usepackage{graphicx}
\usepackage{subcaption}

\title[Validating future gravity missions via optical clock networks]
  {Potential and scientific requirements of optical clock networks for validating satellite gravity missions}
\author[Stefan Schröder et al.]
  {Stefan Schröder$^1$, Simon Stellmer$^2$, Jürgen Kusche$^1$ \\
  $^1$ Institute of Geodesy and Geoinformation, University of Bonn\\
  $^2$ Physics Institute, University of Bonn\\
  }
\date{}
\pagerange{\pageref{firstpage}--\pageref{lastpage}}
\volume{}
\pubyear{}

\begin{document}

\label{firstpage}

\maketitle

\begin{summary}
The GRACE and GRACE-FO missions have provided an unprecedented quantification of large-scale changes in the water cycle. Equipped with ultraprecise intersatellite ranging links, these missions collect measurements that are processed further to monthly sets of spherical harmonic coefficients of the geopotential, which then can be converted into estimates of surface mass variation. However, it is still an open problem of how these data sets can be referenced to a ground truth.\\
Meanwhile, stationary optical clocks show fractional instabilities below 10$^{-18}$ when averaged over an hour, and continue to be improved in terms of precision and accuracy, uptime, and transportability. The frequency of a clock is affected by the gravitational redshift, and thus depends on the local geopotential; a relative frequency change of 10$^{-18}$ corresponds to a geoid height change of about $1$ cm. Here we suggest that this effect could be further exploited for sensing large-scale temporal geopotential changes via a network of clocks distributed at the Earth's surface. In fact, several projects have already proposed to create an ensemble of optical clocks connected across Europe via optical fibre links.\\
Our hypothesis is that a clock network spread over Europe -- for which the physical infrastructure is already partly in place -- would enable us to determine temporal variations of the Earth's gravity field at time scales of days and beyond, and thus provide a new means for validating satellite missions such as GRACE-FO or a future gravity mission.
Mass changes at the surface of an elastic Earth are accompanied by load-induced height changes, and clocks are sensitive to non-loading (e.g. tectonic) height changes as well. As a result, local and global mass redistribution and local height changes will be entangled in clock readings, and we argue that very precise GNSS height measurements will be required to separate them.\\
Here, we show through simulations how ice (glacier mass imbalance), hydrology (water storage) and atmosphere (dry and wet air mass) variations over Europe could be observed with clock comparisons in a future network that, however, follows current design concepts in the metrology community. We assume different scenarios for clock and GNSS uncertainties and find that even under conservative assumptions -- a clock error of $10^{-18}$ and vertical height control error of $1.4$ mm for daily measurements -- hydrological signals at the annual time scale and atmospheric signals down to the weekly time scale could be observed.
\end{summary}


\section{Introduction}
Launched in 2002, the Gravity Recovery and Climate Experiment (GRACE) has observed the time-variable gravity field of the Earth for over fifteen years. GRACE data have opened new avenues in fields such as the terrestrial water cycle, sea level rise, and glacier research (e.g. \citealp{tapley_contributions_2019, wcrp_global_sea_level_budget_group_global_2018}).
Since 2018, the GRACE data record is continued by the successor mission GRACE-FO \citep{flechtner_what_2016}, and several strategies for Next Generation Gravity Missions (NGGMs) have been proposed. Improved or new instruments like the GRACE-FO Laser Ranging Interferometer and different orbit configurations (e.g. \citealp{panet_earth_2013, elsaka_comparing_2014}) promise to further improve the uncertainty and resolution of the gravity fields.
\par
While the GRACE/-FO data are widely used now in environmental monitoring and starts to enter Earth system model simulations via data assimilation \citep{schumacher_systematic_2016}, it is still difficult to assess the 'true' errors of these data products. Thus, several techniques have been proposed for validation of GRACE data. Satellite Laser Ranging (SLR) has been used to derive gravity field variations for a long time, but only low degrees can be utilized for validating GRACE spherical harmonic coefficients \citep{cheng_unexpected_2017}. Using superconducting gravimeters (SG) for validating GRACE is under debate due to their sensitivity to local hydrological conditions and wet air masses (see \citealp{van_camp_quest_2014} and \citealp{crossley_comment_2014}). Also, several studies identified a clear common signal between GNSS derived height changes and GRACE \citep{davis_climate-driven_2004, chanard_toward_2018}. However, tectonics, glacial isostatic rebound, aquifer compaction and groundwater recharge all affect GNSS measurements of deformation but do not follow elastic loading theory and cannot be derived from GRACE measurements only. This is critical in Europe in particular, where the loading signal is comparably low.
In addition, it has been recently proposed to declare Terrestrial Water Storage (TWSA) an Essential Climate Variable (ECV), which will further underline the significance of monitoring TWSA, as well as providing validation for satellite data products.
\par
Optical clocks \citep{ludlow_optical_2015} have seen a steady increase in performance over the past 30 years. Today's best clocks in leading metrology laboratories reach fractional instabilities at or below the $10^{-18}$ range, and are continuously improved in terms of stability and accuracy, dead-time free interrogation \citep{schioppo_ultrastable_2017}, transportability \citep{koller_transportable_2017}, and continuous realization of a timescale \citep{grebing_realization_2016}.
By comparing the tick-rates of two resting clocks, one can measure the difference in gravity potential acting on these clocks, due to the relativistic redshift effect. A frequency shift of one part in $10^{18}$ corresponds to about $0.1$ m$^2$ s$^{-2}$ potential difference, or $0.01$ m geoid height difference.
Clock comparisons over large spatial scales have been conducted in Europe between national metrology institutes Système de Références Temps-Espace (SYRTE) in Paris and Physikalisch-Technische Bundesanstalt (PTB) in Braunschweig \citep{lisdat_clock_2016}. Further terrestrial links have been established between Paris and London as well as between Polish metrology laboratories. In the framework of the project Clock Network Services (CLONETS) it has been suggested to extend these existing links to a pan-European network \citep{krehlik_clonets_2017}.
\par
Terrestrial fibre link technology has evolved over recent years, largely driven by communication needs. However, while all potential network sites considered in this study are already connected to fibre networks, standard infrastructure and common network protocols do not allow for stabilities in the $10^{-16}$ range. Whether fibre networks can achieve stabilities required for optical clock comparison, depends on investment in tailored hardware: \cite{turza_stability_2020} suggest that with duplex unidirectional signal transfer in existing (i.e. operational, soil-deployed) dense wavelength--division multiplexing (DWDM) telecommunication networks one can achieve 1-day averaging stability in the range $10^{-16}$. \cite{chiodo_cascaded_2015} built a coherent fibre link over more than 1000 km with uncertainties of $10^{-16}$ at 1 s integration time and $10^{-20}$ at 60.000 s integration time, via augmenting existing DWDM internet infrastructure by custom-made hardware in a few network nodes (their signal ran largely parallel with the internet data traffic). Using a dedicated fibre network employing cascaded Brillouin amplification finally, \cite{raupach_brillouin_2015} demonstrated optical frequency transfer uncertainties at the ${10^{-20}}$ level and below.
\par
Clock comparisons have been evaluated against measured height differences \citep{takano_geopotential_2016, mehlstaubler_atomic_2018}, but here we go one step further, by suggesting that for validating mass change from GRACE-FO and NGGM's one would have to keep errors below the limits set by high-precision geophysical GNSS monitoring equipment. This means that errors in clock comparisons would have to be corresponding to the millimeter scale or $10^{-19}$ fractional frequency uncertainty. A network of clocks with a corresponding low noise floor would improve over monitoring mass-induced loading with geodetic GNSS networks alone because it would be sensitive to mass redistribution induced potential change as well as elastic loading, but it would also observe non-elastic (e.g. human-induced) and non-loading (e.g. tectonic) height changes. However, in combination with collocated permanent GNSS measurements for height control, the mass effect would be separable. We expect that this combination would be favourable compared to networks of terrestrial gravimeters because the gravity potential is less influenced by local mass changes than the gravity acceleration.
\par
\citet{voigt_time-variable_2016} provide an overview of the relevant tidal and non-tidal time-variable effects that are expected to alter clock rates. \citet{lesmann_analysis_2018} analysed the effect of non-tidal ocean loading at hypothetical German and French clock locations, concluding that potential variations can reach amplitudes of up to $0.5$ m$^2$ s$^{-2}$ at coastal sites ($5\times 10^{-17}$). Along a different line of reasoning, \citet{lion_determination_2017} investigated the use of optical clocks in combination with measured gravity disturbances for regional geoid determination.
\par
In this study, we propose to assess, via simulations, how non-tidal time-variable geophysical signals could be observed with a network of clocks, equipped with collocated, high-precision GNSS height control. We construct three noise scenarios, assuming white noise for the clock fractional frequency readings with standard deviations of $10^{-18}$ ($10^{-19}$, $10^{-20}$) and a combination of flicker and white noise for GNSS with errors corresponding to a total noise variance of $1.4$ ($0.7$, $0.3$) mm. We neglect noise contributions stemming from the fibre link, as experimental studies have shown that this contributes less than $10^{-20}$ when dedicated fibre hardware is used \citep{raupach_brillouin_2015, lisdat_clock_2016}. We simulate time series of clock comparisons and subsequently analyse these in the time and frequency domain, which then enables us to identify which geophysical signals could be detected under which scenario. Despite the fact that geophysical mass change signals are low compared to other regions of the world, here we focus on Europe since infrastructure is already existing and the likelihood of establishing a clock network with the required noise floor in the coming decade is comparably high. We leave out tidal effects and non-tidal ocean mass and loading effects, and focus on atmosphere, hydrology, and glaciers.
\par
This paper is structured as follows. In Section 2, we provide a background to chronometric geodesy and describe how we simulate noise of GRACE, clocks, and GNSS. Section 3 covers our simulated clock network and the signal simulation over Europe, which is then analysed in the spectral domain in Section 4. We discuss implications of our results in Section 5, and we discuss a few issues that will occur with the realization of a clock network. In Section 6 we summarize our conclusions.

\section{Background and noise modelling}

\subsection{Background}
%\todo{ich wuerde mit kursiv sparsam umgehen, insbesondere in Ueberschriften. Was willst du mit \emph{and noise} oben aussagen? Dass andere das typisch vergessen? Geht es um Geophysical noise? oder um data noise?}
Large-scale hydrological, climatic and geophysical processes are accompanied by mass redistribution at the Earth's surface, which in turn cause geopotential changes $\delta V$, gravity changes, and orbit perturbations of satellites. The GRACE and GRACE-FO missions measure these perturbations via inter-satellite ranging and precise orbit determination, and subsequent level 1 - 2 processing provides the geopotential with monthly resolution.
It is this signal that we propose to be identified in clock comparisons here, and that can be represented through the dimensionless, fully normalized spherical harmonic coefficients (SHC) $\delta v_{nm}$
%\todo{$\delta V$ nach Eq. (1) ist relevant weil DAS das Geophyical Signal aus der kapitelueberschrift ist, und das was man mit GRACE/FO misst - die Zeilgroesse - und was wir hier aus den Frequenzvergleichen bekommen wollen. Das sollte zu Beginn dieses Abschnittes aber auch gesagt werden - ich vermisse das hier}
\begin{align}
\delta V = \frac{GM}{a} \sum_n \left(\frac{a}{r}\right)^{n+1} \sum_m \delta v_{nm} \bar{Y}_{nm}
\label{eq:v_from_stokes}
\end{align}
where $\delta v_{nm}$ is a vector of the Stokes coefficients $s_{nm}$ and $c_{nm}$, and the spherical harmonic functions $\bar{Y}_{nm}$ contain the Legendre functions. 
$GM$ is the geocentric gravitational constant, $a$ the semimajor axis of the Earth taken as a reference radius, and $r$ the radial distance to the geocenter. In what follows we will always consider mass changes at the surface of the Earth and thus identify $r$ with $a$, but this does not limit the applicability of our proposed approach. Under the thin-shell assumption, Eq. (\ref{eq:v_from_stokes}) can be easily solved for the spherical harmonic coefficients of a surface mass distribution that generates $\delta V$. Taking further into account the elastic loading of the Earth's crust, the surface mass change field described by the SHC's $\delta \mu_{nm}$ is related as follows (e.g. \citealp{wahr_time_1998}):
\begin{align}
\delta v_{nm} = \frac{1+k'_n}{2n+1} \delta \mu_{nm}.
\end{align}
The dimensionless load Love numbers $k'_n$ characterize the response of the Earth's crust to the mass loading. With this the geopotential ($\delta V$), vertical surface displacement ($\delta h$) corresponding to elastic loading theory and the effective geopotential ($\delta V - g\delta h$) at the Earth's surface as defined in \citet{voigt_time-variable_2016} would be, respectively:
\begin{align}
\delta V = \frac{GM}{a} \sum_{n=0}^{\infty} \sum_{m=0}^n \frac{1+k'_n}{2n+1} \delta \mu_{nm} \bar{Y}_{nm},
\end{align}
\begin{align}
\delta h = \frac{GM}{a} \sum_{n=0}^{\infty} \sum_{m=0}^n \frac{h'_n}{2n+1} \delta \mu_{nm} \bar{Y}_{nm},
\end{align}
\begin{align}
\delta U = \delta V - g\delta h = \frac{GM}{a} \sum_{n=0}^{\infty} \sum_{m=0}^n \frac{1+k'_n-h'_n}{2n+1} \delta \mu_{nm} \bar{Y}_{nm},
\end{align}
Equation (4) introduces the load Love numbers $h'_n$, denoting the elastic response of the Earth in terms of vertical surface displacement. For later use, we mention that the $h'_n$ should correspond to $k'_n$ in the sense that both are derived from the same 1D Earth model. Equation (5) is what a clock, sitting on the deforming crust, would measure (where we have set $g=-\frac{GM}{a^2}$).
\par
For a single clock resting on the Earth's surface, temporal changes of the effective potential $U$ would lead to temporal changes in fractional frequency due to gravitational redshift (e.g. \citealp{ludlow_optical_2015}):
\begin{align}
\frac{\delta f}{f} = \frac{\delta U}{c^2},
\end{align}
where $f$ is the atomic resonance frequency, $\delta f$ is the change in the clock's frequency, and $c$ equals the speed of light.
For two clocks resting at different locations with geopotential $U_1$ and $U_2$ we have a difference
%Clock comparison can be used to determine the difference in potential between to resting clocks:\todo{nicht so lehrbuchhaft schreiben. UND: wir sind mit dem Potential begonnen und schreiben jetzt die theoretischen Beziehungen zu den Observablen hin. Daher muesste delta f etc links stehen und U rechts. Wenn es so waere wie du oben schreibst clock .. can be used to determine .. dann ist das der Analyseprozess von Messungen. Das ist was anderes, da würde man eben auch fragen muss ich filtern, etcetc. und die Groessen waeren mit Fehlern behaftet! Da hatte ich dich in meiner Email drum gefragt. Die resultierende Formel würde ich dann unter den folgenden Abschnitt schreiben}
\begin{align}
\frac{\Delta f}{f} = \frac{\Delta U}{c^2} = \frac{U_2 - U_1}{c^2}.
\end{align}
In the following we will look at the temporal changes of the effective potential $\delta U$ that will occur for a terrestrial clock due to geodynamic and environmental/climate effects. We are interested in the sub-mm to cm level in terms of the geoid, which translates into $10^{-3}$ to $10^{-1}$ m$^2$ s$^{-2}$ of geopotential and into $10^{-20}$ to $10^{-18}$ fractional frequency.
We assume that such uncertainty could be reached through averaging clock readings over periods of several minutes to a day.
A comparison, when implemented over longer time and with a certain temporal resolution (i.e. a finite averaging interval) would then enable one to determine the time-variable contribution in $\Delta U$:
\begin{align}
    \delta U_2 - \delta U_1 = \frac{(\delta f_2 + \varepsilon_{\delta f_2}) - (\delta f_1 + \varepsilon_{\delta f_1})}{f} \cdot c^2
\end{align}
with the two clocks' respective errors $\varepsilon_{\delta f}$.
In order to retrieve the time series of differential geopotential $\delta V$ at a fixed reference height instead of differential effective geopotential $\delta U$, one would need to correct for the vertical uplift $\delta h$:
\begin{align}
    \delta V_2 - \delta V_1 &= (\delta U_2 + g \delta h_2) - (\delta U_1 + g \delta h_1)\\
    &= \left( \frac{\delta f_2 + \varepsilon_{\delta f_2}}{f} \cdot c^2 + g (\delta h_2 + \varepsilon_{\delta h_2})\right) - \left( \frac{\delta f_1 + \varepsilon_{\delta f_1}}{f} \cdot c^2 + g (\delta h_1 + \varepsilon_{\delta h_1})\right),
\label{eq:V_U}
\end{align}
where we would have to account for errors in height corrections $\varepsilon_{\delta h}$.
% here would probably come the further part from Jürgen's notes about all effects considered in GRACE/GNSS measurements

% \begin{itemize}
% \item Simulating the signal: time-variable geopotential difference for two clock locations: spherical harmonic representation of mass change (from model or GRACE or both), corresponding geopotential change and fractional shift, indirect effect and vertical loading; what GRACE etc. observe and what GNSS / gravimeters observe
% \item same for tides (?)
% \item copy Eqs. from separate tex file
% \item discussion of spatial and temporal scale where these Eqs. apply and of what is omitted
% \item discussion of the omitted part (geophysical noise)
% \end{itemize}

\subsection{GRACE/-FO noise}
Due to the missions' orbital pattern and the anisotropic sensitivity of the inter-satellite leader-follower configuration, spherical harmonic coefficients from GRACE/-FO typically at monthly resolution are corrupted by anisotropic, latitude- and degree-dependent, and temporally non-stationary noise.
This noise is induced by a combination of instrument errors, temporal aliasing of insufficiently corrected short-term mass signals, and orbit errors. Even nearly two decades after the GRACE launch, it is still difficult to realistically characterize errors for level-2 products, i.e. harmonic coefficients.
This is mainly since no other measurement technique is able to determine geoid height variations with superior accuracy at large spatial scales. In lieu, comparing level-2 data from different processing centers has been suggested, but since processing strategies and background models are similar this cannot be expected to provide a realistic error level. Fitting simple models of trend and annual/semi-annual signal to GRACE data and estimating data errors from such fits has been suggested, but we know from geophysical modelling that real mass signals are rich in episodic and interannual variability which renders this approach overly conservative. Instead of a (still missing) community-agreed GRACE error model, we here use formal errors from \citet{mayer-gurr_itsg-grace2018_2018}.
We propagate these formal errors to geoid height errors at two assumed clock locations, in order to derive geopotential error variances and covariances at these locations. Consecutively, the (co-)variances are propagated to the variance of geoid height difference between the locations -- this procedure enables us to take into account the effect of geographical latitude due to GRACE orbit convergence as well as the distance between clock locations, since larger-scale-errors tend to cancel out for differences at shorter distances.
\par
GRACE-FO errors have been estimated as somewhat lower than for GRACE, due to the more precise laser ranging instrument \citep{flechtner_what_2016}. For future gravity missions, simulation studies have predicted errors as low as ten to twenty times below GRACE-FO noise depending on orbit configuration, number of satellites and instrument types, but this will be likely limited by errors inherent to the short-time background geophysical models. In summary, we assume that future missions will surpass GRACE-FO, but that methods for assessing their errors will be dearly required.
%\begin{itemize}
%\item \todo{Erstmal sagen was GRACE/FO in Bezug auf das vorher eingeführte ueberhapt messen (sollen) - das zeitintegral ueber die $u_{nm}$ fuer 1 Monat minus dr Mittelwert}
 %   \item Expected / propagated errors\todo{kurze diskussion warum es schwierig ist tats. Sigmas anzugeben, wie man es grunds. machen kann - (1) versch. Loesungen vergleichen (2) Trend/Sinus Fit und Residuen rechnen (3) Fehlerfortpflanzung = formal errors - und nichts davon optimal ist - und das wir formal errors nehmen weils wichtig ist die volle Cov zu haben}
    %\item ITSG-2018 \citep{kvas_itsg-grace2018_2019} formal errors used \todo{klaeren wie robust das ist vergl. mit anderen - ausserdem wuerde ich das ans ende dieses subkapitel schreiben, immer erst das allgemeine}
    %\item these formal errors in spherical harmonic form are expanded to monthly, point-wise geoid height errors at the clock locations (eq. (\ref{eq:u_from_stokes})), shtools
    %\item comparison of the GRACE geoid height error of the difference between two clock locations:
%\end{itemize}
%\todo{Die Formel ist mA nach Inkorrekt. Die Varianz einer Potentialdifferenz ist nicht die Summe der Varianzen auf den beiden Punkten, weil das Potential aus GRACE doch raeumlich korreliert ist. Oder?}
%\begin{itemize}
%    \item with $\sigma_{cx}$ being the ITSG-GRACE formal error at the location of clock x
%\end{itemize}
%\todo{Immer erst allgemein formulieren - being the formal error of the satellite gravity solution \dots}

\subsection{Clock network noise}
Clock comparisons yield effective potential differences $\Delta U$, possibly at daily resolution (e.g. \citealp{chou_optical_2010, ludlow_optical_2015}. A clock's uncertainty is usually characterized by an error resulting from systematic uncertainties and an error originating in the instability, expressed through its Allan deviation \citep{riley_handbook_2008}. By definition, any error related to the instability averages out with longer measurement time. For many state-of-the-art optical clocks the instability averages down below the systematic uncertainty within several minutes to hours \citep{bothwell_jila_2019, oelker_demonstration_2019}, albeit with our application of daily, relative measurements in mind, we are not interested in the systematic errors. 
What limits the noise floor here are effects like black body radiation or dc Stark shift, which need to be monitored permanently and for every clock individually \citep{ludlow_optical_2015, bothwell_jila_2019}. 
Because the properties of the errors stemming from this monitoring are not well investigated, we deem it sufficient here to simulate errors as white noise.
% By systematic we do not mean a frequency offset that is constant in time, but a systematic relation of cause and effect, where the cause (e.g. black body radiation, dc Stark shift, etc.) needs to be monitored permanently and for every clock individually \citep{ludlow_optical_2015, bothwell_jila_2019}. Because the properties of the errors stemming from this monitoring are not well investigated, we deem it sufficient here to simulate errors as white noise.
\par
To account for technological progress, we will adopt three different scenarios for clock uncertainties, see Table \ref{tab:scenarios}. In scenario 1, here we work with a noise floor of $10^{-18}$, which is what current state-of-the-art clocks are just reaching now \citep{bothwell_jila_2019, oelker_demonstration_2019}; even contemporary transportable clocks approach this order of magnitude \citep{takamoto_test_2020}. In scenario 2 we assume that clock noise can be decimated by one order of magnitude ($10^{-19}$) and this is what we expect clocks to reach in five to ten years. Scenario 3, in which we assume noise again decreased by one magnitude, i.e. $10^{-20}$, is a "best case" scenario for the clock network. It should be mentioned that these error assumptions refer to single clocks. The error of a clock comparison originates from the noise of the two clocks and the error that the fibre link adds; we will thus assume $1.4 \times 10^{-18} (10^{-19}, 10^{-20})$ uncertainty for a clock comparison. Experimental studies have shown that the fibre link itself contributes less than $10^{-20}$ to the clock comparison \citep{raupach_brillouin_2015}, and we thus neglect noise contributions from the fibre link in our studies, independent of the distance.
% We neglect the uncertainty that is added by the fibre link itself because we expect it to not add significant noise, as it is the case today for dedicated fibre hardware \citep{raupach_brillouin_2015, lisdat_clock_2016}. \todo{clarify that this means we assume dedicated fibre hardware (\dots) and that this is beyond software solutions on existing network or freespace links as with satellite}
\begin{table}
\caption{Clock network and colocated GNSS error scenarios.}
\begin{tabular}{lrrr}
\hline 
\textbf{Error type} & \textbf{Scenario 1} & \textbf{Scenario 2} & \textbf{Scenario 3}\\
\hline 
Clock  white noise& $10^{-18}$ & $10^{-19}$ & $10^{-20}$\\
GNSS white noise & $1$ mm & $0.5$ mm & $0.2$ mm\\
GNSS flicker noise & $4.4$ mm/yr$^{0.25}$ & $3.1$ mm/yr$^{0.25}$ & $2.0$ mm/yr$^{0.25}$\\
\hline 
\end{tabular}
\label{tab:scenarios}
\end{table}
\par
As has been suggested in the above, in order to separate the potential $\delta V$ from the effective potential $\delta U$ we need to correct for $g\delta h$ (Eq. \ref{eq:V_U}). It is obvious that errors in $\delta h$ fully translate into geoid height error; 1 mm height or geoid height error corresponds to $1.1\times10^{-19}$ fractional frequency error. In line with the GNSS literature\citep{williams_error_2004, klos_combined_2017}, we assume that vertical displacement time series derived from GNSS are affected by both white noise and flicker noise, where the latter can be identified with power-law noise with spectral index $\kappa=-1$. White noise in GNSS errors results e.g. from phase errors or unmodelled short term propagation effects, flicker noise from tide model or orbit constellation errors \citep{agnew_finding_2007}. In the noise model that we construct here, we assume that both white noise and flicker noise contribute evenly to the total variance in the daily measurements we simulate. The values for the GNSS flicker noise in Table \ref{tab:scenarios} are computed after \citet{bos_fast_2013}.
Orientating towards \citet{klos_error_2016} and \citet{gruszczynska_multichannel_2018} we 
assume that white noise and flicker noise add each about $1$ mm$^2$ noise variance, resulting in an assumed overall height error Root Mean Square (RMS) of $1.4$ mm. This corresponds to assuming that each clock is accompanied by an IGS-type (International GNSS Service, \citealp{villiger_international_2020}) GNSS antenna and that time-variable height differences between clock and local antenna can be controlled with superior accuracy, e.g. by repeated levelling.
\par
Under scenario 2 and 3 we assume that both white noise and flicker noise will be reduced. We hypothesize that under scenario 2, i.e. within five to ten years, noise in troposphere and GNSS orbit and clock errors etc. can be further reduced, e.g. through rapid development of multi-GNSS analyses. Under scenario 3, we assume that errors can be -- as for the clock rate uncertainties -- further reduced to a level of $0.3$ mm, which would likely require next-generation GNSS systems \citep{giorgi_advanced_2019, glaser_reference_2020}. The assumed reduction of uncertainty between the scenarios is larger for the clocks than for GNSS, which is probably realistic considering the evolution in clock techniques over the last decades \citep{poli_optical_2013}, while GNSS for height control will be always limited by errors in tropospheric and ionospheric modelling. In addition, GNSS could in future benefit from common-clock linking to the National Metrology Institutes' optical clock time reference.
\par
At this point we notice that the typical RMS fits from GRACE-derived vertical deformation or from geophysical load models evaluated versus time series of GNSS height measurements can still be a few millimeters \citep{chanard_toward_2018} -- as these comparisons employ elastic loading theory and detrending, these fits must inevitably include residual vertical motion signals as well as troposphere errors and site-specific effects. As a result, such comparisons cannot serve for developing a GNSS error model.
\par
We expect that for comparing monthly geopotential change derived from GRACE/-FO or from future gravity missions and the daily, relative fractional frequency changes within a clock network, one would convert the fractional frequency changes to geoid height changes and average these to monthly data. We do not expect that the clock network errors will average out following a white noise behaviour, i.e. decrease with averaging time. Rather we expect that certain error contributions will be reduced, like errors in tide models, whereas others such as clock drift may increase. In the absence of a better knowledge of the individual error sources, we will assume here that monthly averages of relative clock rates will not further improve over daily rates. 
For the computation of monthly GNSS error estimates we take temporal correlations into account by using the Power Spectral Density (PSD) of the power-law noise that we constructed; the integral over the PSD, starting at monthly frequencies, equals the variance of monthly errors.

\section{Simulation Setup}
\subsection{Network}
\begin{table}
\caption{Stations assumed here to participate in a clock comparison network. The national metrology institutes are written in bold.}
\begin{tabular}{llllr}
\hline 
\textbf{Location} & \textbf{Institution} & \textbf{Abbr.} & \textbf{Country} & \textbf{Lon/Lat [$^\circ$]}\\
\hline 
Vienna & Federal Office of Metrology and Surveying & \textbf{BEV} & Austria & 16.4/48.2\\
Prague & CESNET Association of legal entities & CESNET & Czech Republic & 14.5/50.1\\
Helsinki & National Metrology Institute of Finland & \textbf{MIKES} & Finland & 24.8/60.2\\
Paris & Time Space Reference Systems, Paris Observatory & \textbf{SYRTE} & France & 2.2/48.8\\
Strasbourg & University of Strasbourg & US & France & 7.8/48.6\\
Braunschweig & Physikalisch-Technische Bundesanstalt & \textbf{PTB} & Germany & 10.4/52.3\\
Bonn & University of Bonn & UB & Germany & 7.1/50.7\\
Munich & Max Planck Institute of Quantum Optics & MPG & Germany & 11.5/48.2\\
Potsdam & GeoForschungsZentrum & GFZ & Germany & 13.1/52.4\\
Wettzell & Geodetic Observatory & GOW & Germany & 12.9/49.1\\
Torino & Italian National Metrology Institute & \textbf{INRiM} & Italy & 7.6/45.0\\
%\textit{Vilnius} & \textit{Center for Physical Sciences and Technology} & \textit{FTMC} & %\textit{Lithuania} & not yet in toolbox\\
Delft & Dutch Metrology Institute & \textbf{VSL} & Netherlands & 4.4/52.0\\
Poznan & Supercomputing and Networking Center & PSNC & Poland & 17.0/52.4\\
Torun & University of Torun & UT & Poland & 18.6/53.0\\
Warsaw & Central Office of Measures & \textbf{GUM} & Poland & 21.0/52.2\\
Ljubljana & Slovenian Institute of Quality and Metrology & \textbf{SIQ} & Slowenia & 14.5/46.0\\
Gothenburg & National Laboratory for Length and Dimensional& \textbf{RISE} & Sweden & 12.0/57.7\\
& Metrology, Research Institutes of Sweden & & & \\
Bern & Federal Institute of Metrology & \textbf{METAS} & Switzerland & 7.4/46.9\\
London & National Physical Laboratory & \textbf{NPL} & United Kingdom & 0.3/51.4\\
\hline 
\end{tabular}
\label{tab:stations}
\end{table}

\begin{figure}
\includegraphics[width=0.8\textwidth]{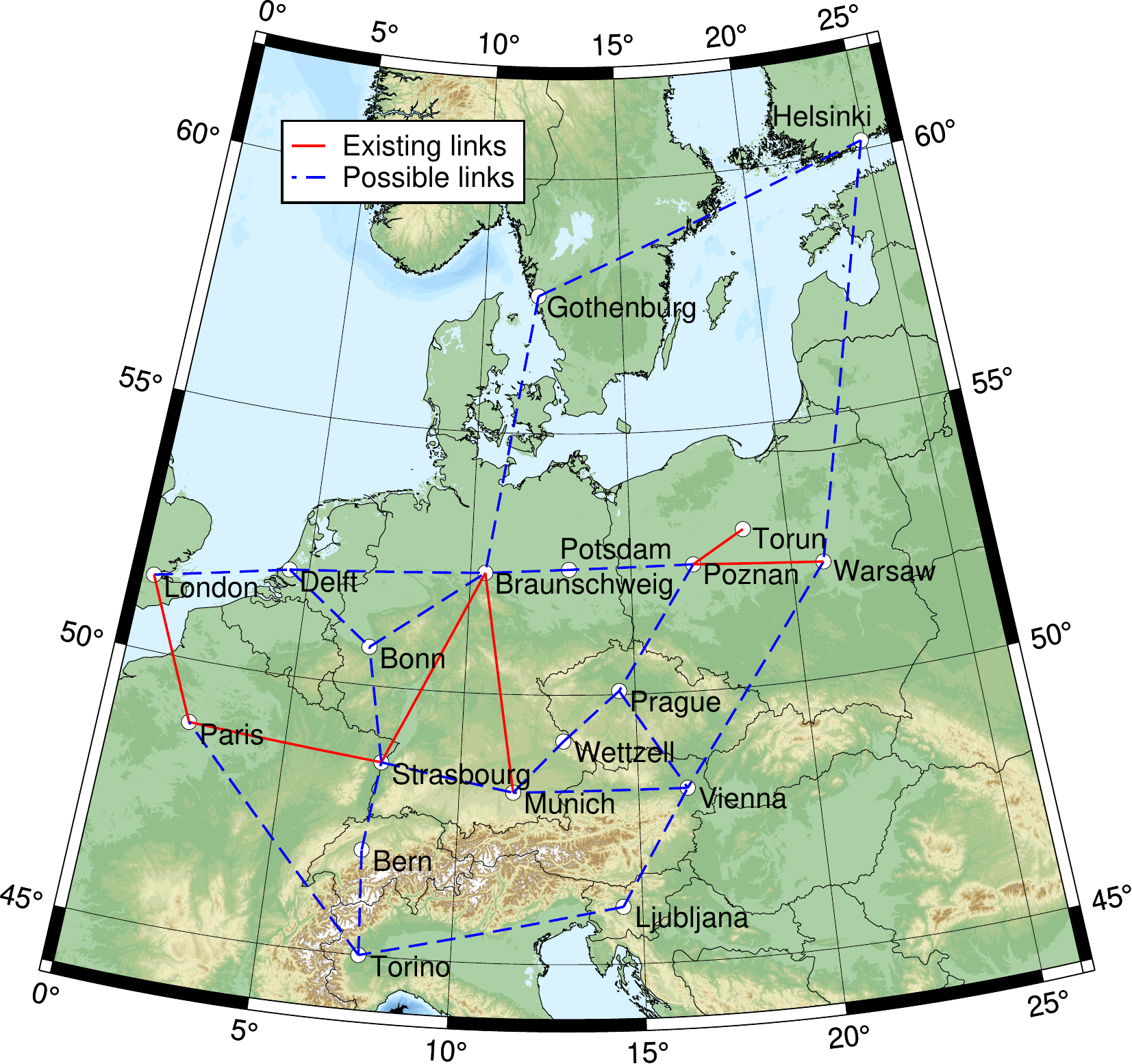}
\caption{Assumed network geometry. Red fibre links with dedicated hardware for ultra-precise clock comparisons are already existing, blue dashed links are additional connections that we consider here.}
\label{fig:network}
\end{figure}

For our simulation we assume that most of Europe's national metrology institutes and other UTC-laboratories will be equipped with ultra-precise clocks, and that fibre links exist already or will be successively established between them. The geographical distribution of the institutes are visualized in Fig. \ref{fig:network} and are as in Table \ref{tab:stations}. Moreover, we took the liberty to add a few additional sites (Wettzell, Potsdam, Bonn) for potential deployment of precise clocks and expansion of the fibre network, which are under consideration for additional metrological experiments in the geodetic context. 
With this setup we take the following assumptions: 1) One optical clock is located at every site marked in Fig. \ref{fig:network}. 2) At every clock, an IGS-type GNSS-site is colocated in order to correct for land elevation change. 3) One would be able to carry out measurements of relative fractional frequency differences between all the clocks during a single session every day, which have an integration time sufficiently long for the instability to drop below the systematic uncertainty.
\par
The outcome of this configuration would be one daily geopotential time series per clock relative to a chosen reference clock. In other words, we assume that one of the clocks would serve as a reference clock for all others. We are aware that other concepts could be devised, e.g. defining a reference frequency standard via averaging all clocks, but this would not change our experiment significantly. In either case, as in GNSS network analyses, large-scale, common-mode geopotential and height change signals that affect all clocks in the same way could not be detected. This is, however, an imperative consequence of the clock comparison method, unless we can anchor clock comparisons via free-space links to a reference clock in space \citep{mehlstaubler_atomic_2018}. In what follows, we explore to what extent such a setup would enable one to detect spatiotemporal changes of the gravity field over Europe, and thus could provide a reference for gravity missions. We do not discuss the objective of realizing a unified height reference system here \citep{lisdat_clock_2016}.
\par
Then, in an additional experiment we extend this setup by assuming that all the EUREF permanent network GNSS stations would be accompanied by clock measurements.
The purpose of this scenario is to understand how beneficial a denser clock network could be for monitoring gravity change and mass redistribution at a spatial scale down to tens of kilometers.

\subsection{Simulated signal}
For a single location, we can write the effective potential as
\begin{align}
U = \bar{U} + \delta U^T + \delta U^{NT} - g \delta h
\end{align}
with time-mean $\bar{U}$, tidal variations $\delta U^T$ and nontidal variations $\delta U^{NT}$. As mentioned before, any vertical motion of the surface $\delta h$ will also translate into a potential change at that surface.
Usually, tidal potential variations are split into direct astronomical (tide-generating) potential $\delta U^{TGP}$, Earth tide potential $\delta U^{ET}$, ocean tides $\delta U^{OT}$, and atmospheric tides $\delta U^{AT}$.
The nontidal potential variation contains contributions from mass redistributions in the atmosphere $\delta U^{A}$, ocean $\delta U^{O}$, cryosphere $\delta U^{I}$, and the solid Earth $\delta U^{S}$, and from Glacial Isostatic Adjustment $\delta U^{GIA}$ and hydrology $\delta U^{H}$, i.e. surface and groundwater storage changes.
\par
Here, we focus on the nontidal part of the potential induced by atmospheric mass variations $\delta U^{A}$ and by storages changes in hydrology $\delta U^{H}$. We simulate atmospheric mass variability following \cite{forootan_comparisons_2013} with with 3D pressure and moisture data from ERA5 \citep{hersbach_global_2019}. $\delta U^{H}$ is derived from water storage in thirty layers simulated with the Community Land Model (CLM; \citealp{oleson_assessment_2003}) 3.5, forced by the Weather Research and Forecasting (WRF; \citealp{skamarock_description_2008}) version V3.3.1. As a result, for these two contributions we obtain daily values of total water storage anomalies. $\delta U^{H}$ and $\delta U^{A}$ are simulated over the year 2007. Mass redistributions due to retreat or thickening of glaciers in the Alps and Scandinavian mountains is simulated with the Open Global Glacier Model (OGGM, \citealt{maussion_open_2019}), which is forced by CRU TS4.01 \citep{harris_updated_2014} data. The temporal resolution of the OGGM output is limited to monthly values, but we construct a longer time series from October 2005 to September 2008. We represent the corresponding potential via spherical harmonic coefficients of maximum degree 720 for $\delta U^{H}$ and $\delta U^{I}$, and degree 180 for $\delta U^{A}$, which corresponds to a half-wavelength of $111$ km and $28$ km, respectively.
\par
\begin{figure}
\begin{subfigure}[c]{0.32\textwidth}
\includegraphics[width=\textwidth]{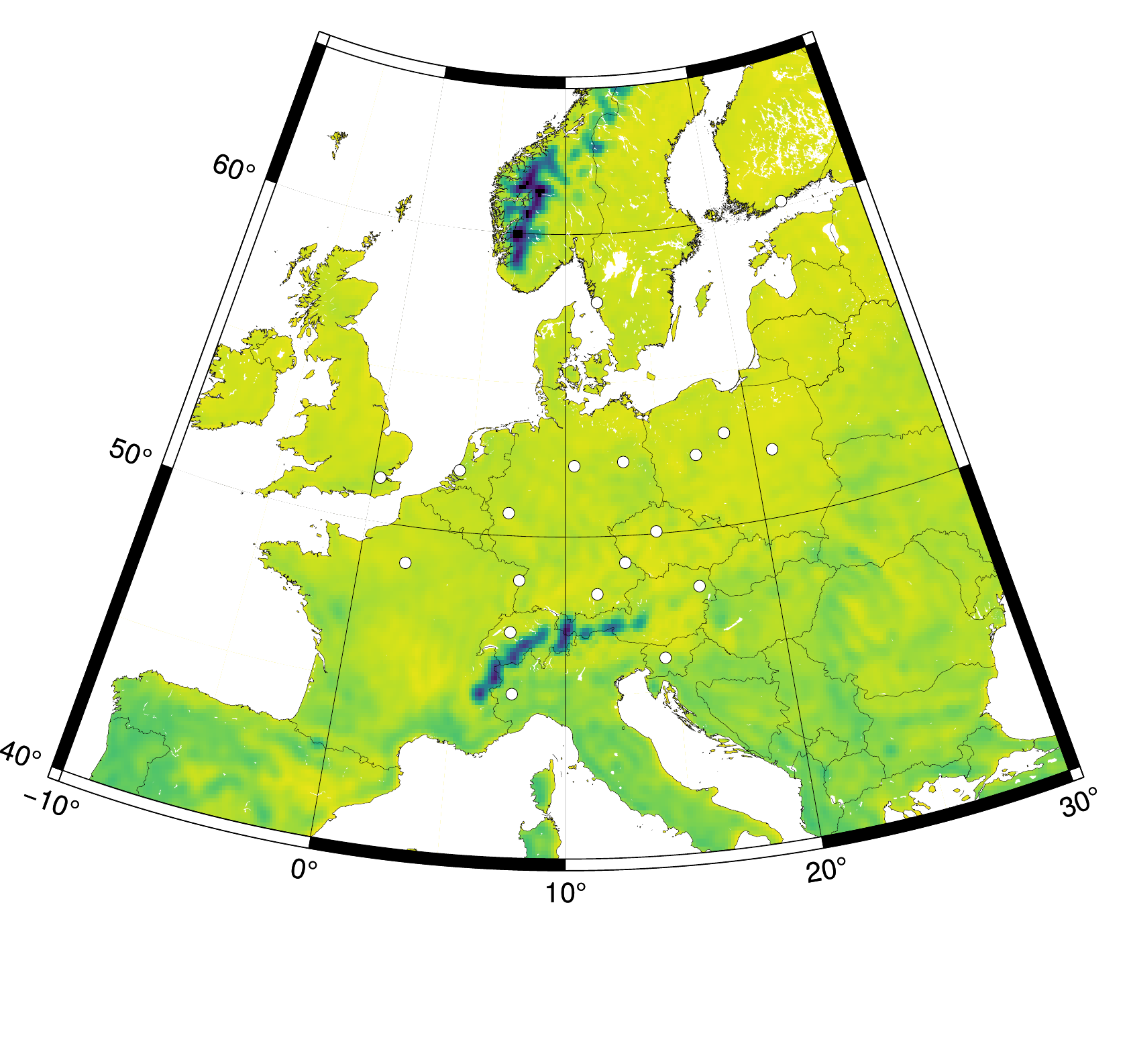}
\end{subfigure}
\begin{subfigure}[c]{0.32\textwidth}
\includegraphics[width=\textwidth]{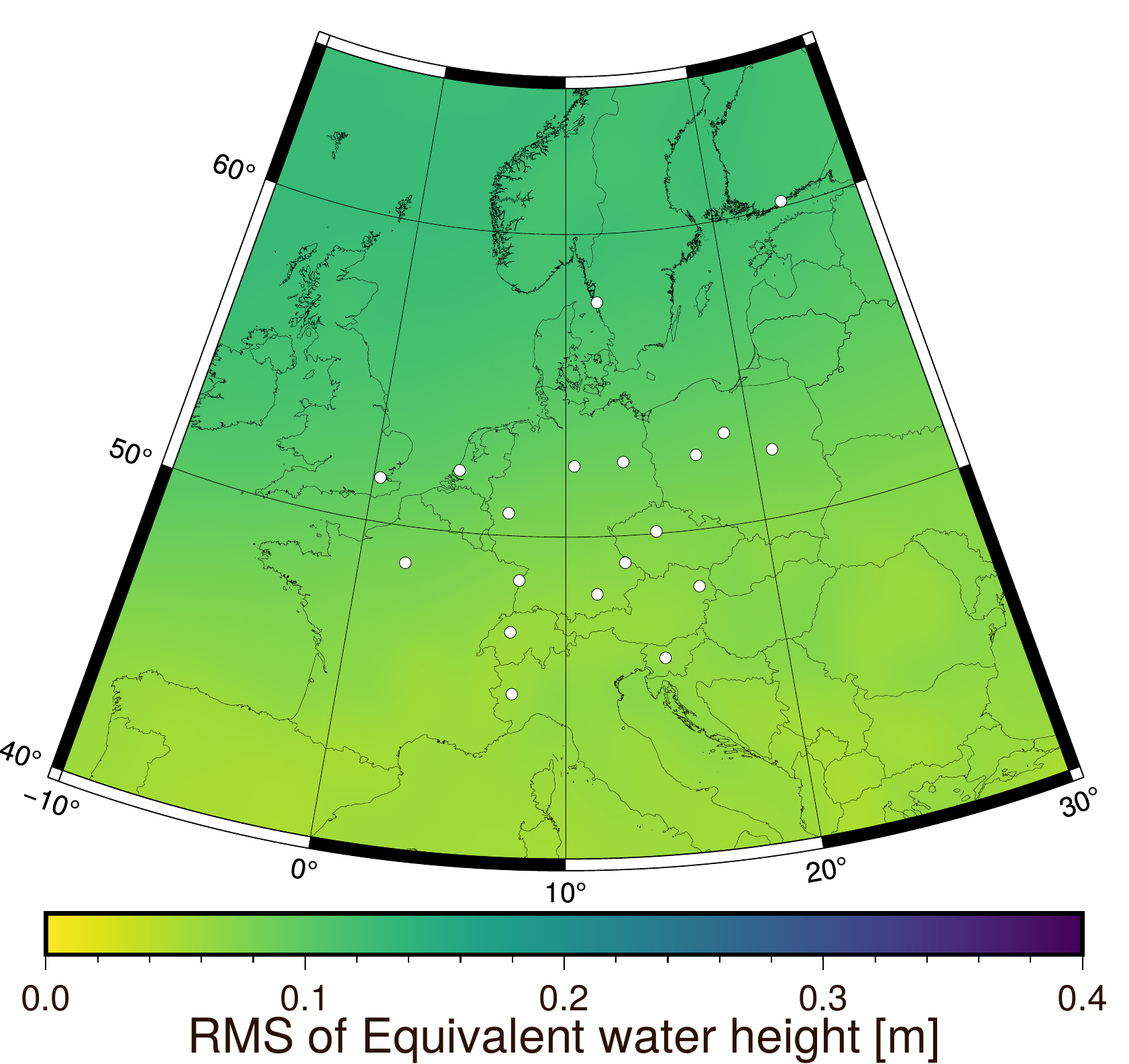}
\end{subfigure}
\begin{subfigure}[c]{0.32\textwidth}
\includegraphics[width=\textwidth]{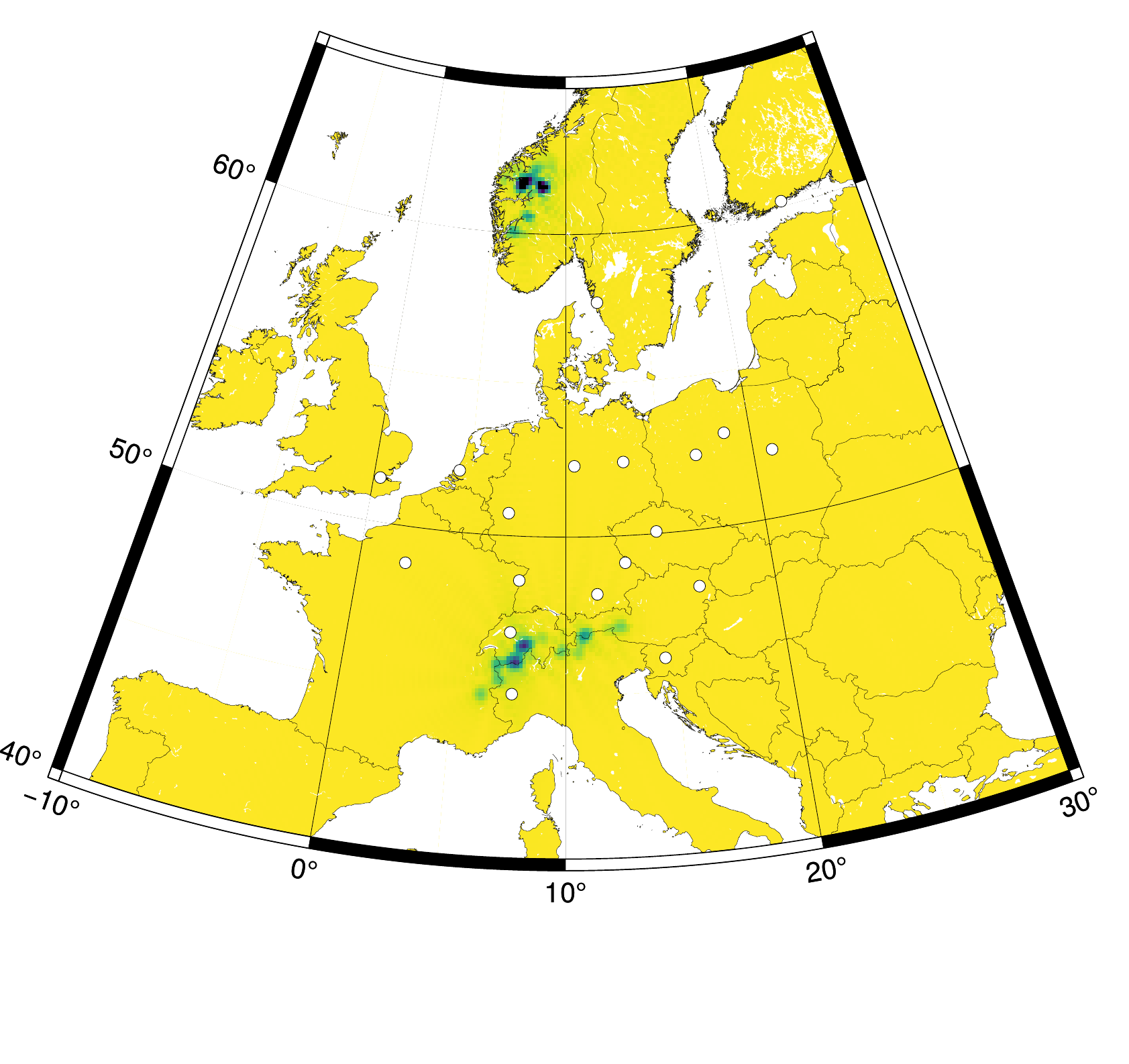}
\end{subfigure}
\begin{subfigure}[c]{0.32\textwidth}
\includegraphics[width=\textwidth]{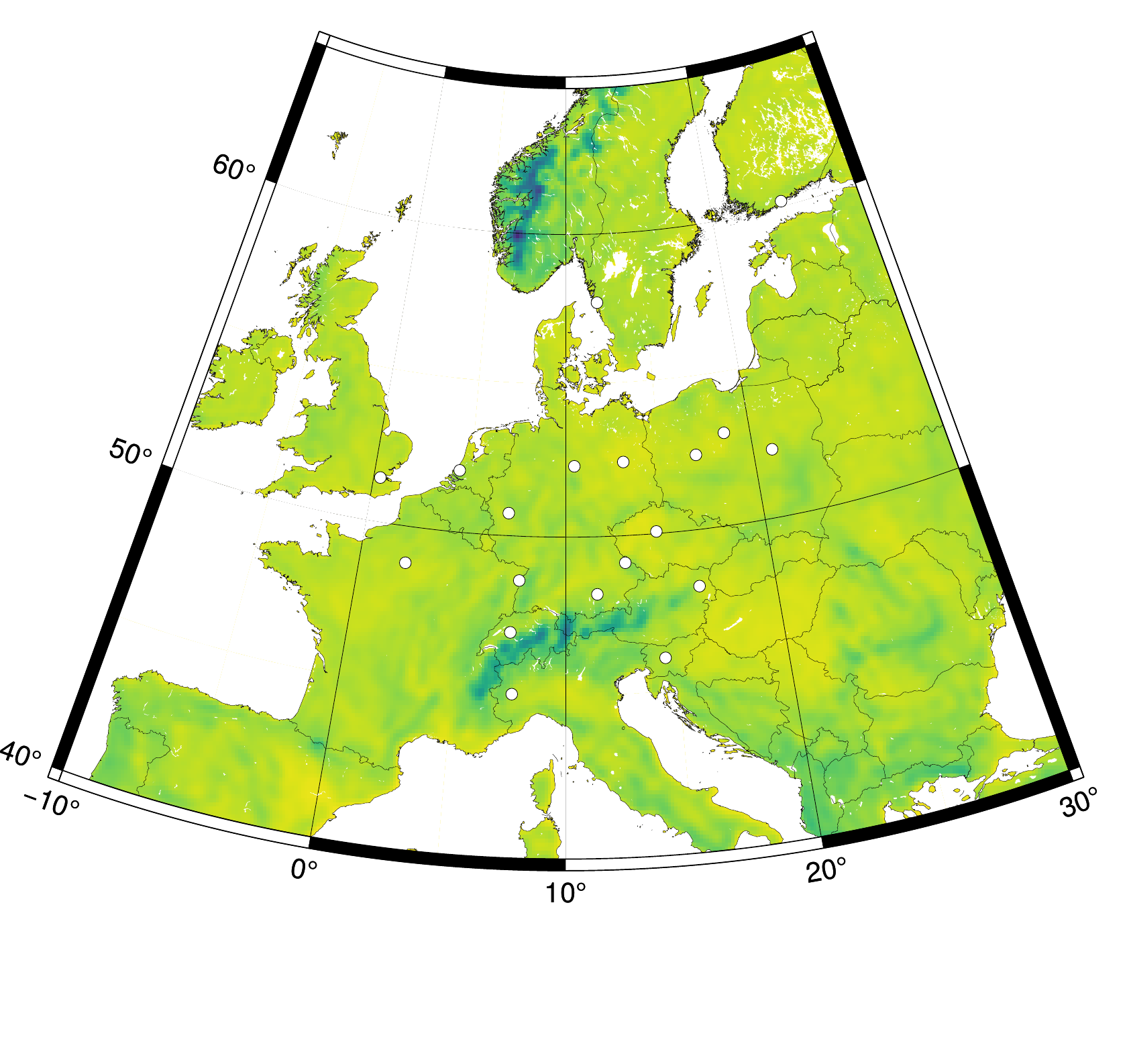}
\end{subfigure}
\begin{subfigure}[c]{0.32\textwidth}
\includegraphics[width=\textwidth]{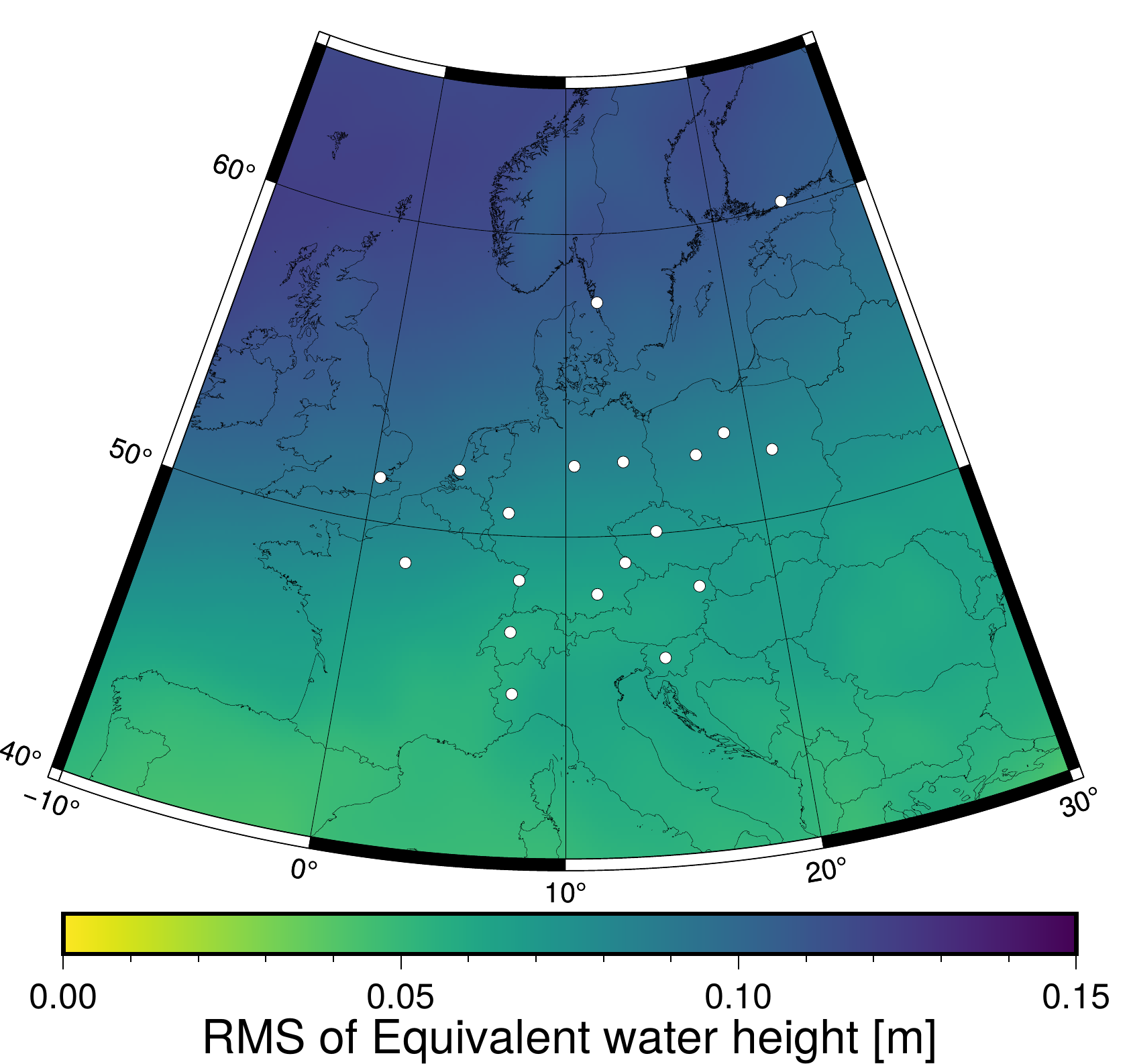}
\end{subfigure}
\begin{subfigure}[c]{0.32\textwidth}
\includegraphics[width=\textwidth]{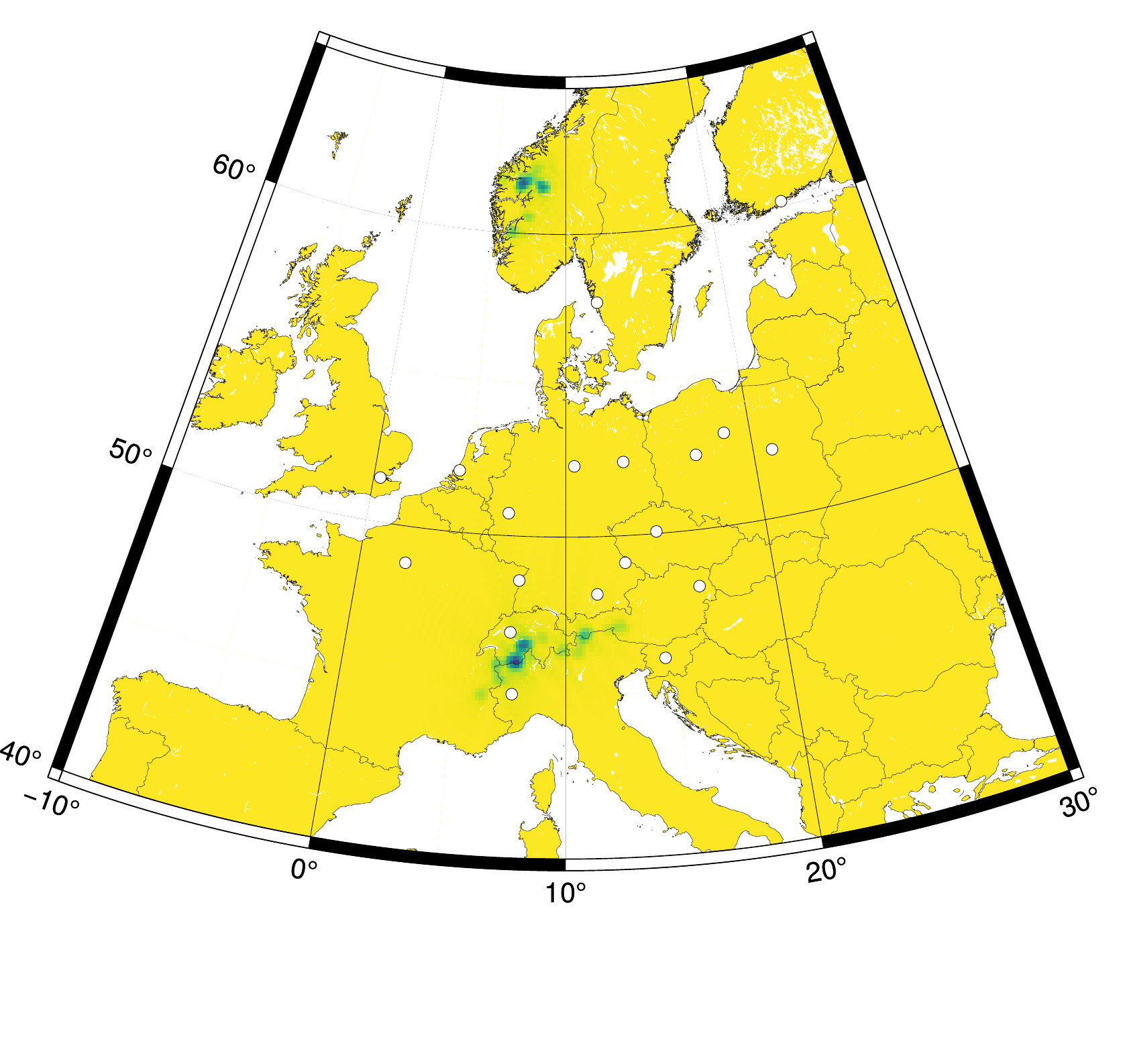}
\end{subfigure}
\begin{subfigure}[c]{0.32\textwidth}
\includegraphics[width=\textwidth]{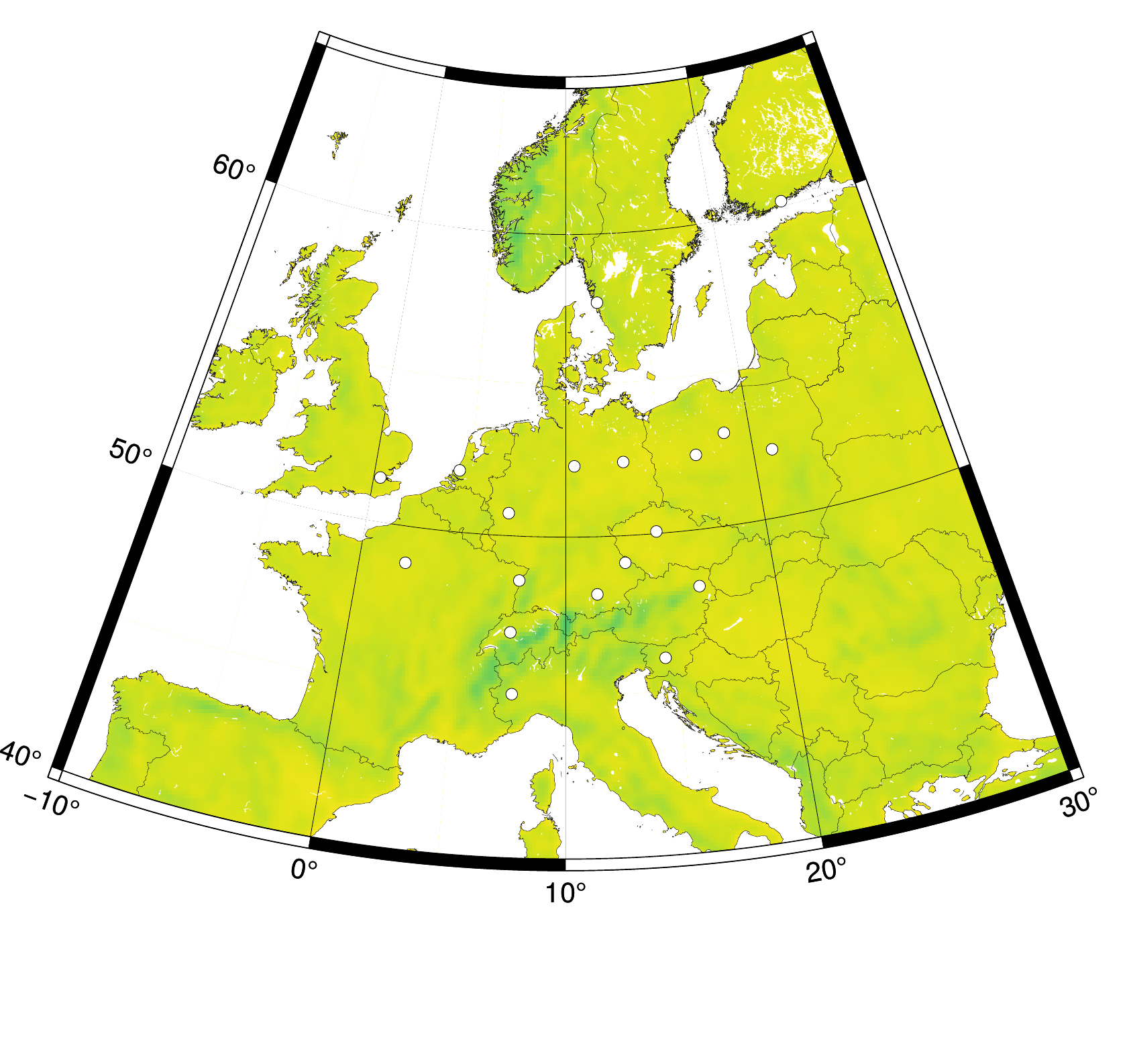}
\end{subfigure}
\begin{subfigure}[c]{0.32\textwidth}
\includegraphics[width=\textwidth]{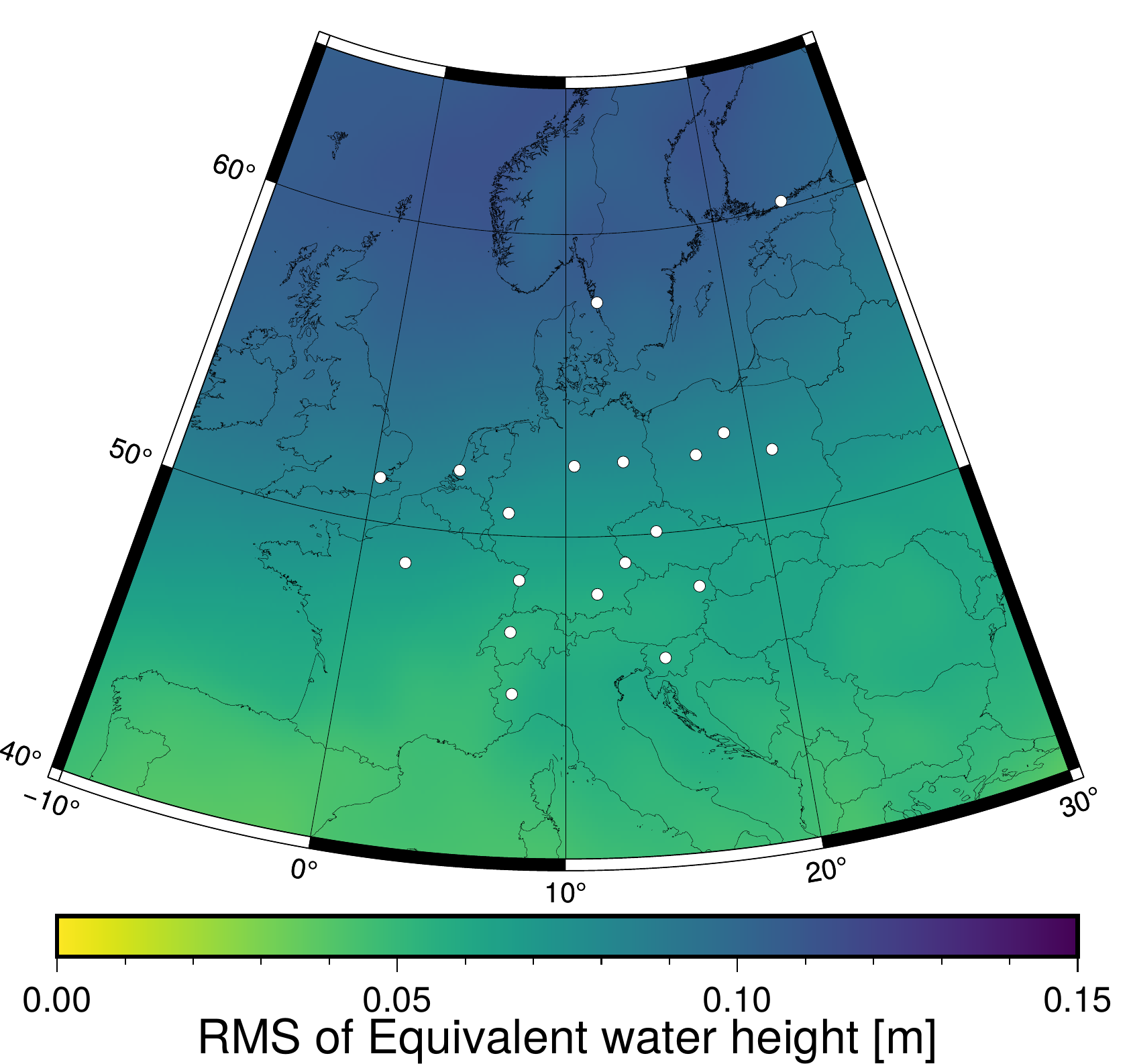}
\end{subfigure}
\begin{subfigure}[c]{0.32\textwidth}
    \hspace{2cm}
\end{subfigure}
\begin{subfigure}[c]{0.32\textwidth}
\includegraphics[width=\textwidth]{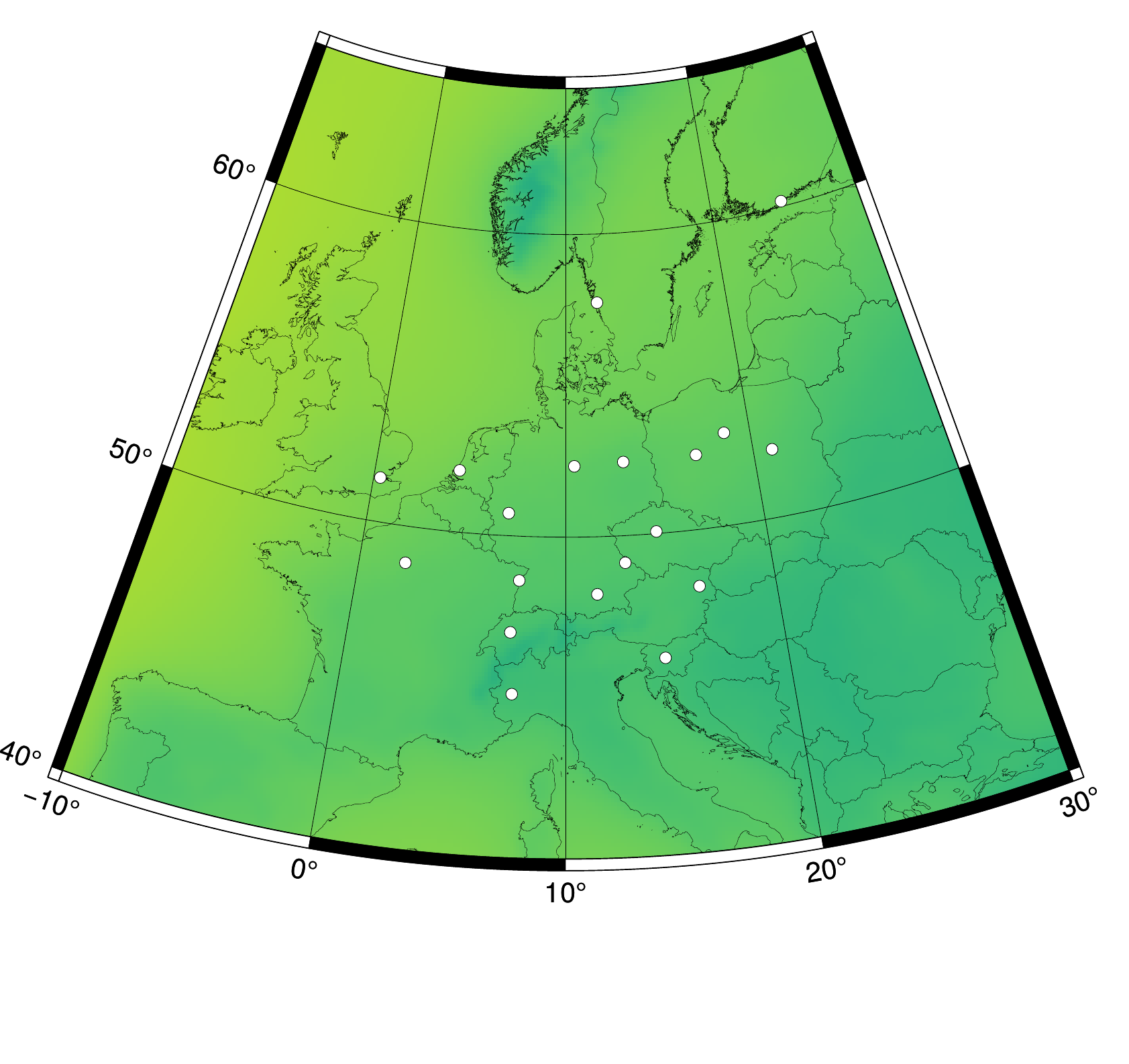}
\end{subfigure}
\hspace{0.18cm}
\begin{subfigure}[c]{0.32\textwidth}
\includegraphics[width=\textwidth]{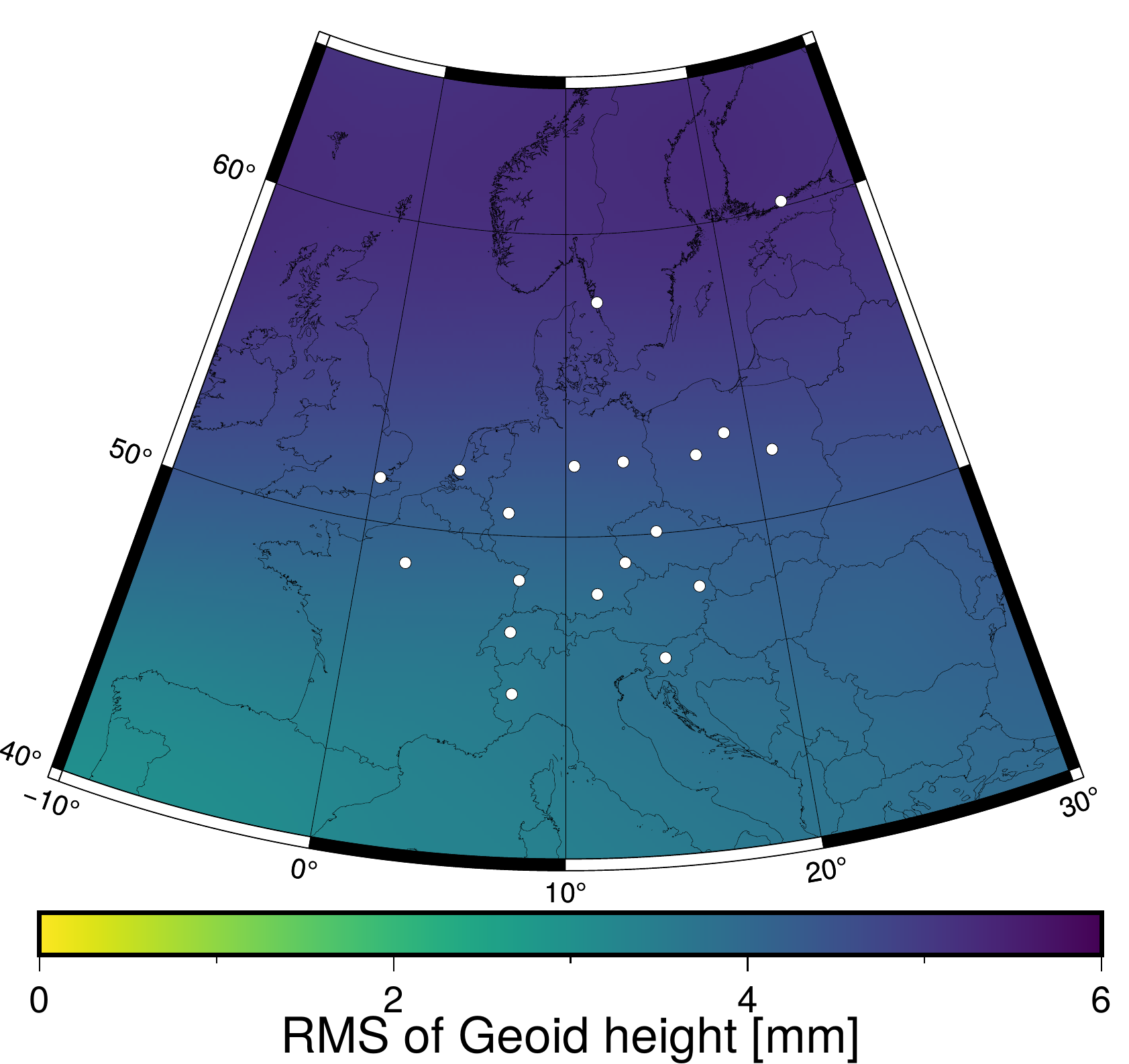}
\end{subfigure}
\hspace{0.18cm}
\begin{subfigure}[c]{0.32\textwidth}
\includegraphics[width=\textwidth]{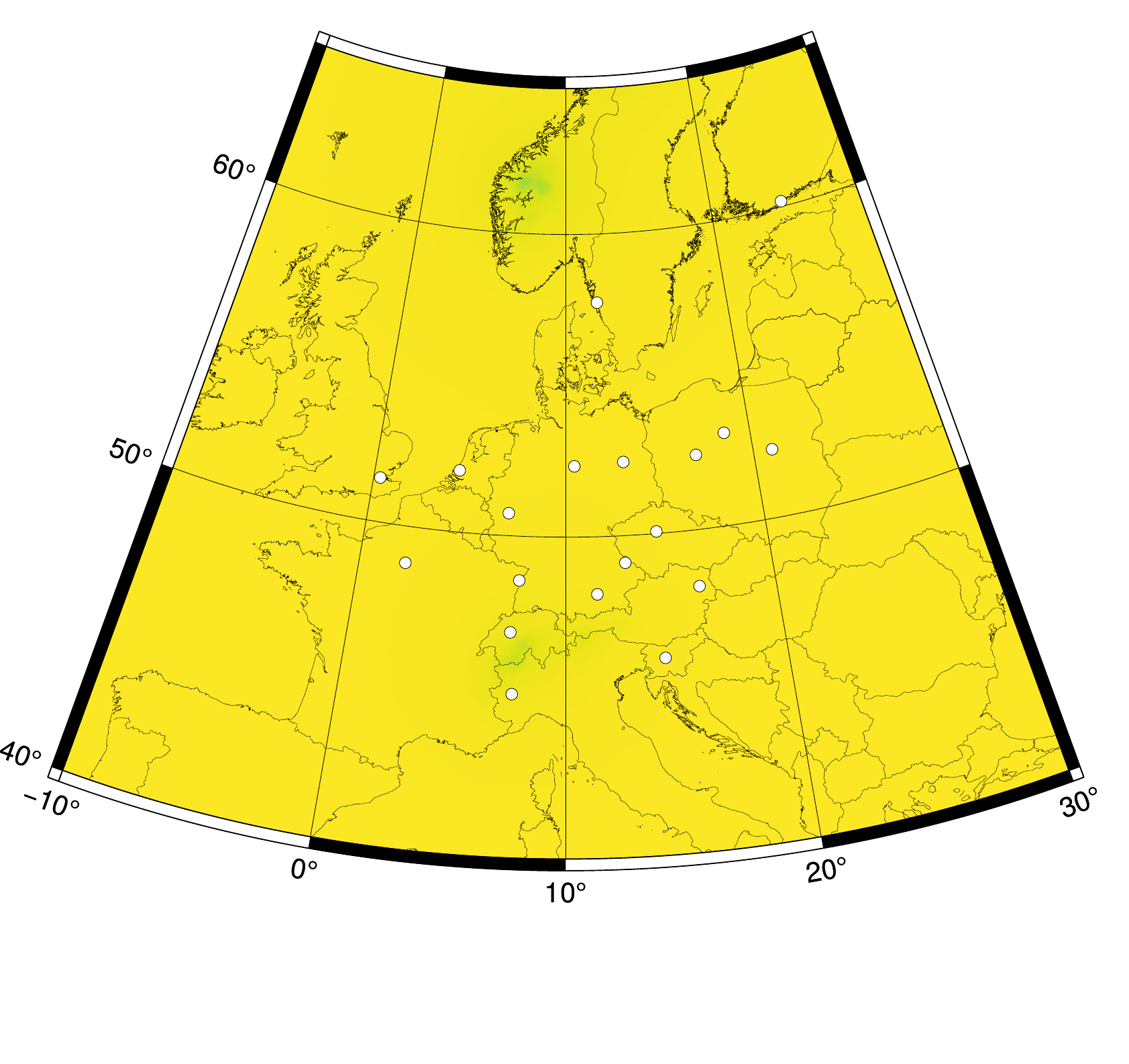}
\end{subfigure}
\caption{2007 equivalent water height root mean square (RMS) variability of daily hydrologic (first column) and atmospheric (second column) mass redistribution signal expressed in equivalent water height, as well as monthly glacial (third column) signal from 2005 to 2008. The first row shows the full signal, the second row shows the RMS after de-trending and de-seasoning, and in the third row the day-to-day signal variability only is shown. The last row shows resulting geoid height RMS of the full signal, thus corresponding to the first row.}
\label{fig:RMS}
\end{figure}

For hydrological, atmospheric, and glacier mass changes, the variability with respect to the 2007 mean (Fig. \ref{fig:RMS}, columns one, two, and three, respectively) reaches up to $40$ ($15$, $40$) cm when expressed in equivalent water height (EWH, first row), and $2$ ($6$, $0.5$) mm in corresponding geoid changes, visualized in the last row.
The hydrological signal shows high variability in the mountain regions of Scandinavia and in the Alps. Since the signal over Europe contains much energy at smaller spatial scales, the effect on the geoid is rather small. This is very different from other regions of the world, such as the Amazon, where geoid variability reaches $1$-$2$ cm.
After removing a six-parameter model that contains linear trend, annual and semiannual signals, the variability drops to $15$ cm EWH; this is shown in the second row of the figure, without corresponding geoid height. 
To focus on the sub-monthly signal, we have additionally applied a thirty-day boxcar filter. This results again in a distinct decrease in variability, reaching about one order of magnitude in the mountain regions compared to the full signal. 
We note that the day-to-day variability is in particular interesting since neither GRACE/-FO nor future gravity missions are likely able to monitor this signal at an appropriate spatial resolution.
\par
For the atmosphere (second column) we see a much smoother picture of EWH variability. This partly stems from the fact that the maximum spherical harmonic degree of our simulated atmospheric potential coefficients is only 180, corresponding to a spatial resolution of 110km. This is governed by the ERA5 spatial resolution, and experiments with high-resolution regional models \citep{dobslaw_modeling_2016} suggest that atmospheric mass variability at higher degrees is predominantly driven by topographic variability and in general low. The main reason for the smoothness though is probably the large size of atmospheric high and low pressure areas. Here, we do not see large differences after removal of long-term signals; in fact the variability is concentrated at higher temporal scales, it is between 11 and 13 cm EWH. This illustrates the low seasonality of the atmospheric mass. For Europe, with up to 5mm geoid height RMS the large scale atmospheric variability transfers to a higher geoid height variability as compared to hydrological storage changes.
\par
The last column shows the RMS variability resulting from glacier simulations in the Alps and the Scandinavian mountains. This shows a picture very similar to that of the hydrology, but somewhat more localized because the variations are limited to glaciated areas. Our hydrological model contains a snow component; it is thus expected that patterns are similar. One might argue that these two models thus simulate similar processes in mountainous regions. However, since the glacier model contains a much more sophisticated representation of ice storage and thus provides a more realistic picture, we have decided to work with all three models. In our analyses we do not interpret the combination of the models in the sense of the sum of effect, though.
\section{Results}
In this section, we analyse time series of simulated fractional frequency variation for individual clock locations, which were simulated using models for atmospheric, hydrological, or cryospheric mass variability over Europe, as explained in Section 3. We are showing exemplary time series for clocks in Braunschweig, London, Bern, Warsaw, Helsinki, and Gothenburg, whose geographical locations are well spread over the network area. Our focus will be on surface mass changes, the resulting vertical displacement, geoid (potential) change, and in particular the fractional frequency change.
% \begin{itemize}
%     \item time series for a few links/clocks comparisons: signal - GRACE-FO error - clock network error
% \end{itemize}

\subsection{Hydrological storage changes}

\begin{figure}
\includegraphics[width=\textwidth]{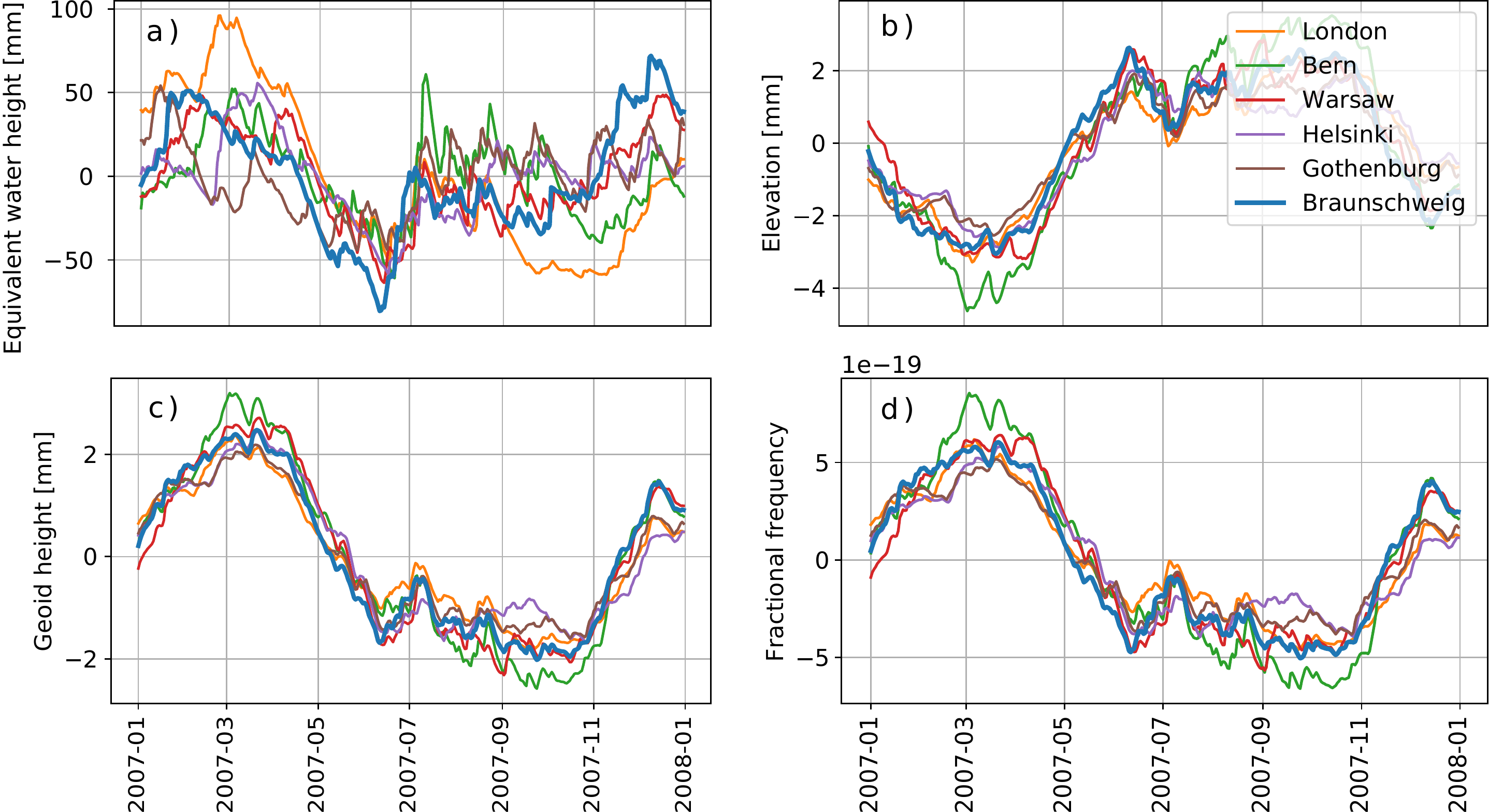}
\caption{a) Daily hydrological storage variations at several stations in 2007 and resulting b) elevation, c) geoid height, and d) fractional frequency variations.}
\label{fig:H}
\end{figure}
% \begin{itemize}
%     \item Fig. \ref{fig:H}
%     \item annual signal very dominant, especially in b) - d)
%     \item much more heterogeneity in equivalent water height variations, because local changes in water storage do not have a large influence on elevation/geoid height
%     \item land elevation and geoid height very similar, just mirrored
%     \item ff follows from these two: well in the range of some 1e-19
%     \item problem: dominant annual signal in all clocks, i.e. not detectable with clock comparisons, see also figures \ref{fig:braunschweig} and \ref{fig:braunschweig_bern}
% \end{itemize}
% \par
Comparing the the time series corresponding to hydrologic storage changes in 2007 (Fig. \ref{fig:H}), the annual signal can be seen in every time series. However, surface mass changes expressed in equivalent water height (EWH, a) differ significantly from clock to clock. This is due to the influence of regional/local hydrology, which is here included up to the 10km scale via the CLM model, i.e. omitting only very local effects. However, such regional-to-local scale mass changes will affect vertical displacement (b) and geoid height change (c) much less than effects of larger spatial scale, thus these two signals are very similar, albeit they show an opposing sign. Simulated displacement and geoid height change for the Bern site exhibit a slightly higher amplitude than at the other stations due to their proximity to the hydrologically highly variable Alps. The fractional frequency (d) variations result directly from the geoid height and the vertical displacement. Although elastic loading theory predicts that land elevates when geoid heights decrease due to surface mass redistribution, they both will contribute to geopotential changes with the same sign, and this will then lead to an amplified fractional frequency change.
It is this fractional frequency change and its magnitude, which is of key interest here. We observe an amplitude of circa $10^{-18}$, thus in the cm-range of physical heights.
\par
% \begin{figure}[ht]
% \includegraphics[width=\textwidth]{figures/H_ff_Braunschweig_both.pdf}
% \caption{a) Time series of hydrology-induceduced fractional frequency variations in \textbf{Braunschweig} and b) the spectral domain of it.}
% \label{fig:braunschweig}
% \end{figure}
\begin{figure}
\begin{subfigure}[c]{0.49\textwidth}
\includegraphics[width=\textwidth]{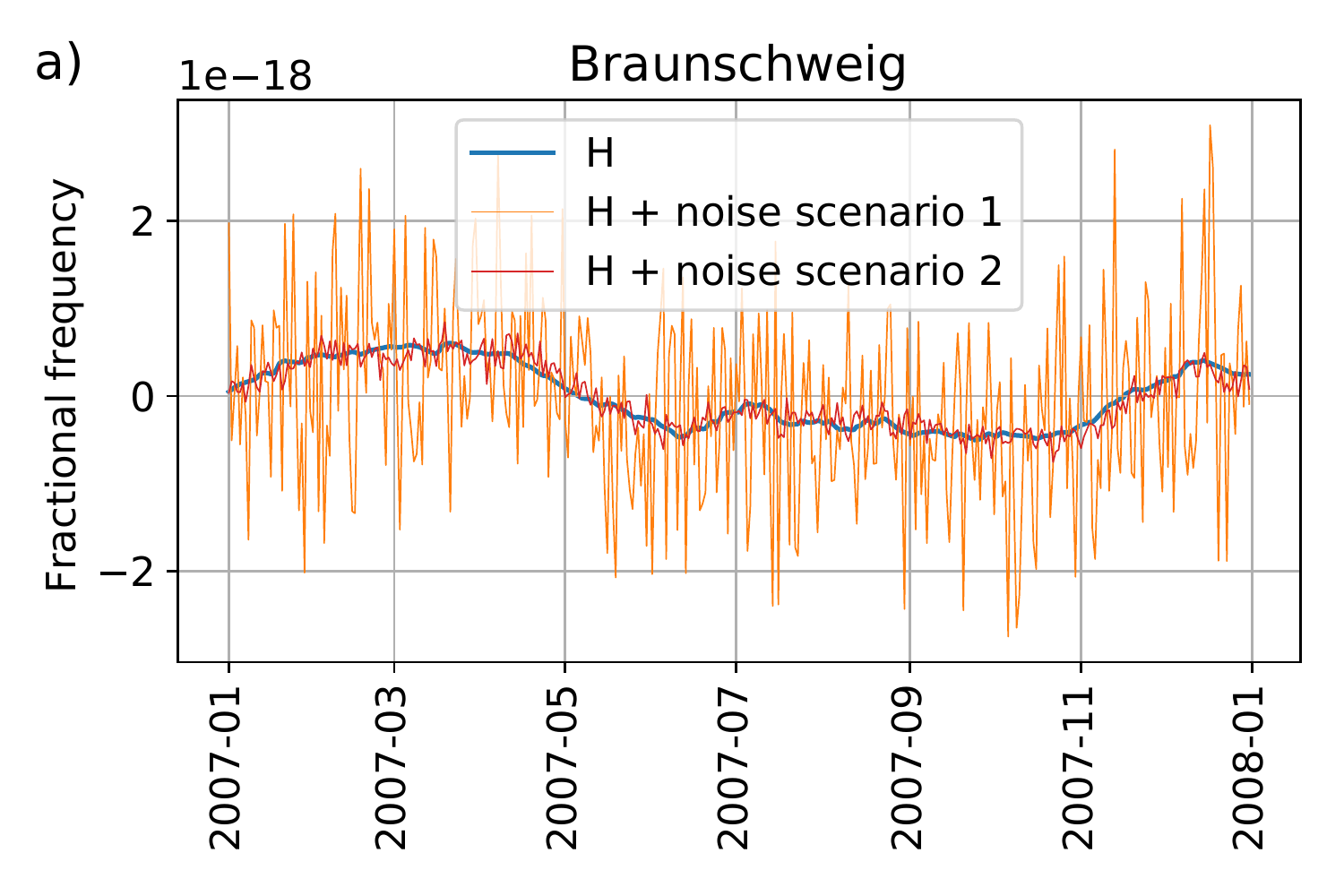}
\end{subfigure}
\begin{subfigure}[c]{0.49\textwidth}
\includegraphics[width=\textwidth]{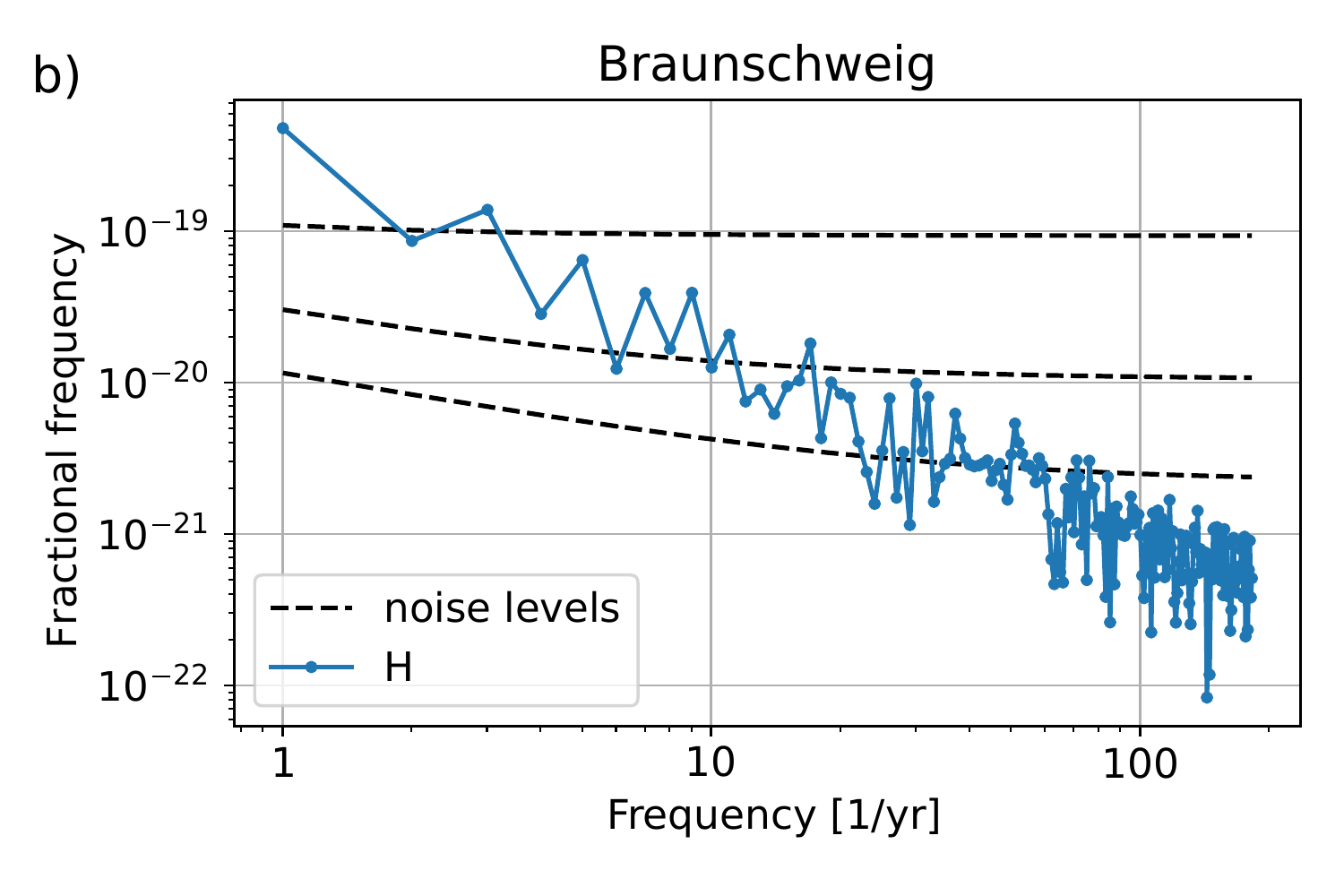}
\end{subfigure}
\caption{a) Comparison time series of hydrology-induced ("H") fractional frequency changes for \textbf{Braunschweig} and b) in the spectral domain.}
\label{fig:braunschweig}
\end{figure}
Exemplarily for Braunschweig, the time series is shown with added noise from scenarios 1 ($10^{-18}$ clock error and $1.4$ mm GNSS error) and 2 ($10^{-19}$ and $0.7$ mm) in Fig. \ref{fig:braunschweig}a). The scenario 3 noise ($10^{-20}$ and $0.4$ mm) is left out since it is too small to be distinguishable. 
The noise levels are also shown in relation to the corresponding amplitude spectrum (\ref{fig:braunschweig}b), where it can be observed that the annual signal is well above the scenario 1 noise level, but to shorter-term signals would not stick out of the noise in this case. In contrast, for scenario 2, frequencies beyond a month would be detectable against the noise floor. We conclude that for identifying signals at even higher frequencies, a lower clock and/or GNSS uncertainty would be needed.
\par
\begin{figure}
\begin{subfigure}[c]{0.49\textwidth}
\includegraphics[width=\textwidth]{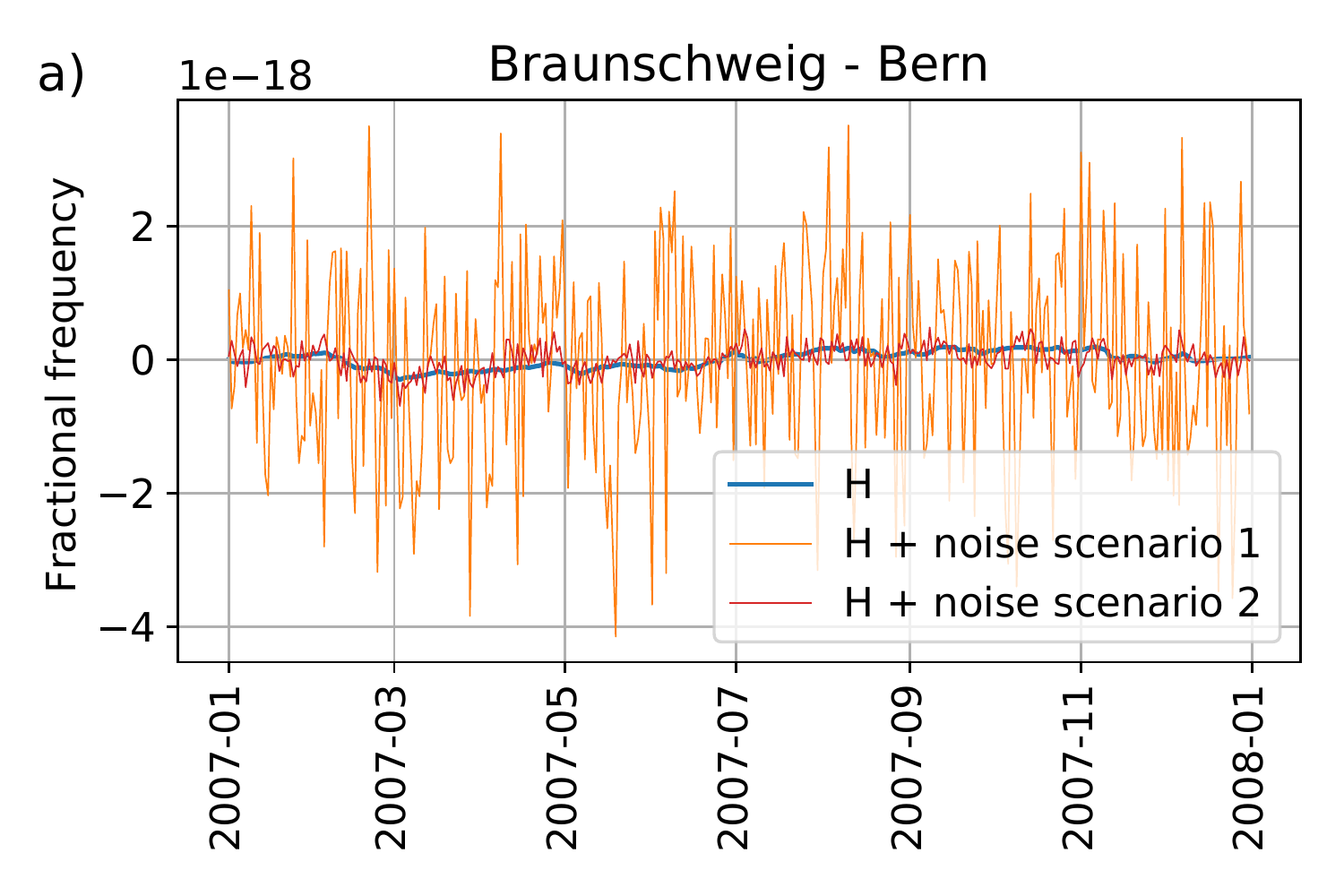}
\end{subfigure}
\begin{subfigure}[c]{0.49\textwidth}
\includegraphics[width=\textwidth]{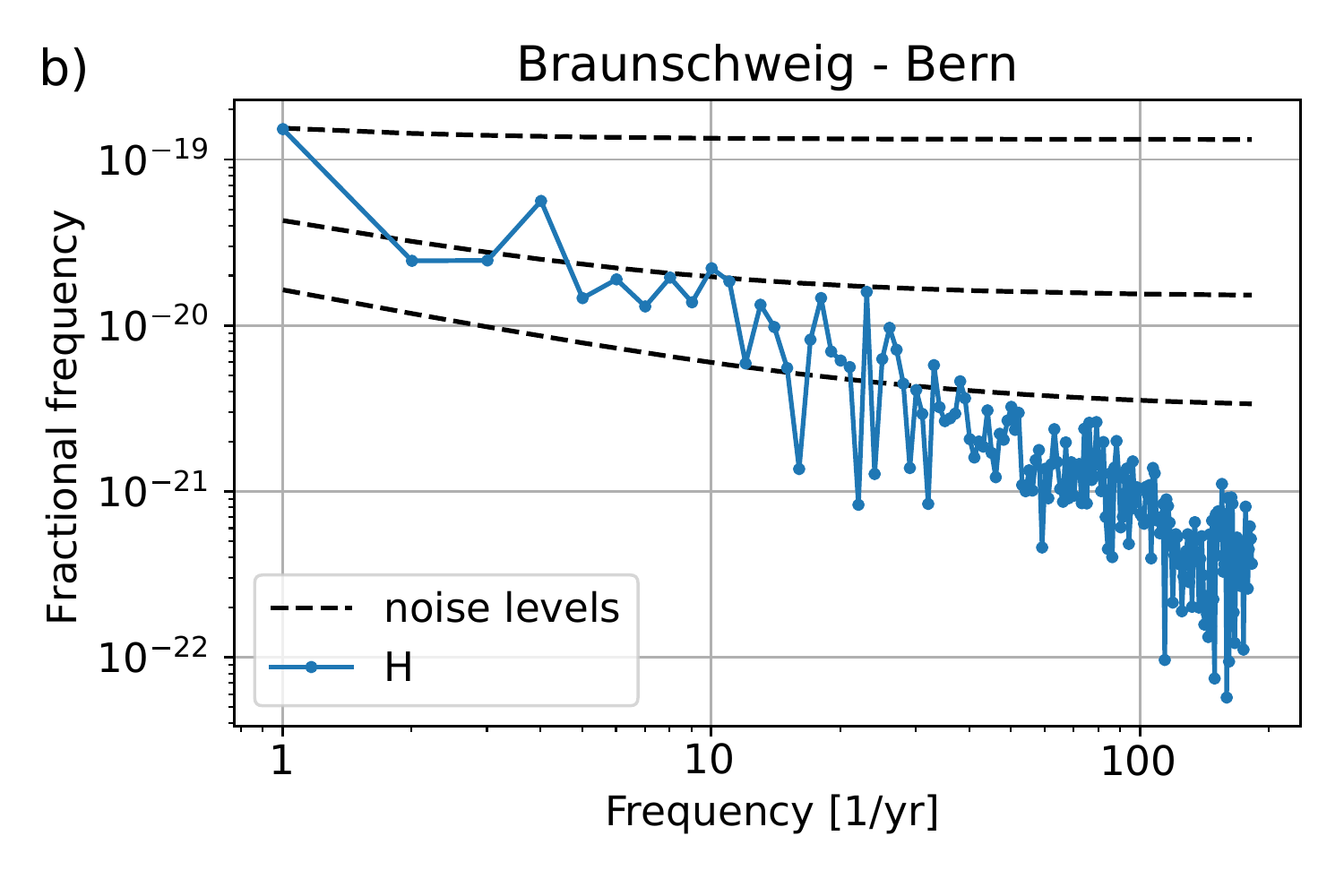}
\end{subfigure}
\caption{a) Comparison time series of hydrology-induced fractional frequency changes for the link \textbf{Bern vs. Braunschweig} and b) the amplitude spectrum of it.}
\label{fig:braunschweig_bern}
\end{figure}
\par
In order to observe changes in clock tick rates, one would need to perform clock comparisons. The simulated clock comparison time series and its corresponding amplitude spectrum between PTB (Braunschweig) and METAS (Bern) can be seen in Fig. \ref{fig:braunschweig_bern}. As expected, the observable relative fractional frequency change would be smaller than for a single clock, because the distance of the two locations is only 630km, thus low degree signals affect both clocks to a similar extent and cancel out. The annual signal is thus reduced from $5\times 10^{-19}$ to $1.5\times 10^{-19}$ and would just reach the noise level of scenario 1, which at the same time is slightly elevated due to the differencing approach. For the higher frequencies, almost no single frequency amplitude would stick out of the noise level of scenario 2, and only signals up to a biweekly frequency are above the assumed noise of scenario 3. This is an important outcome of this analysis, since it suggests that short-term hydrologic mass changes would not significantly affect clock comparisons.
\par
\begin{figure}
\begin{subfigure}[c]{0.49\textwidth}
\includegraphics[width=\textwidth]{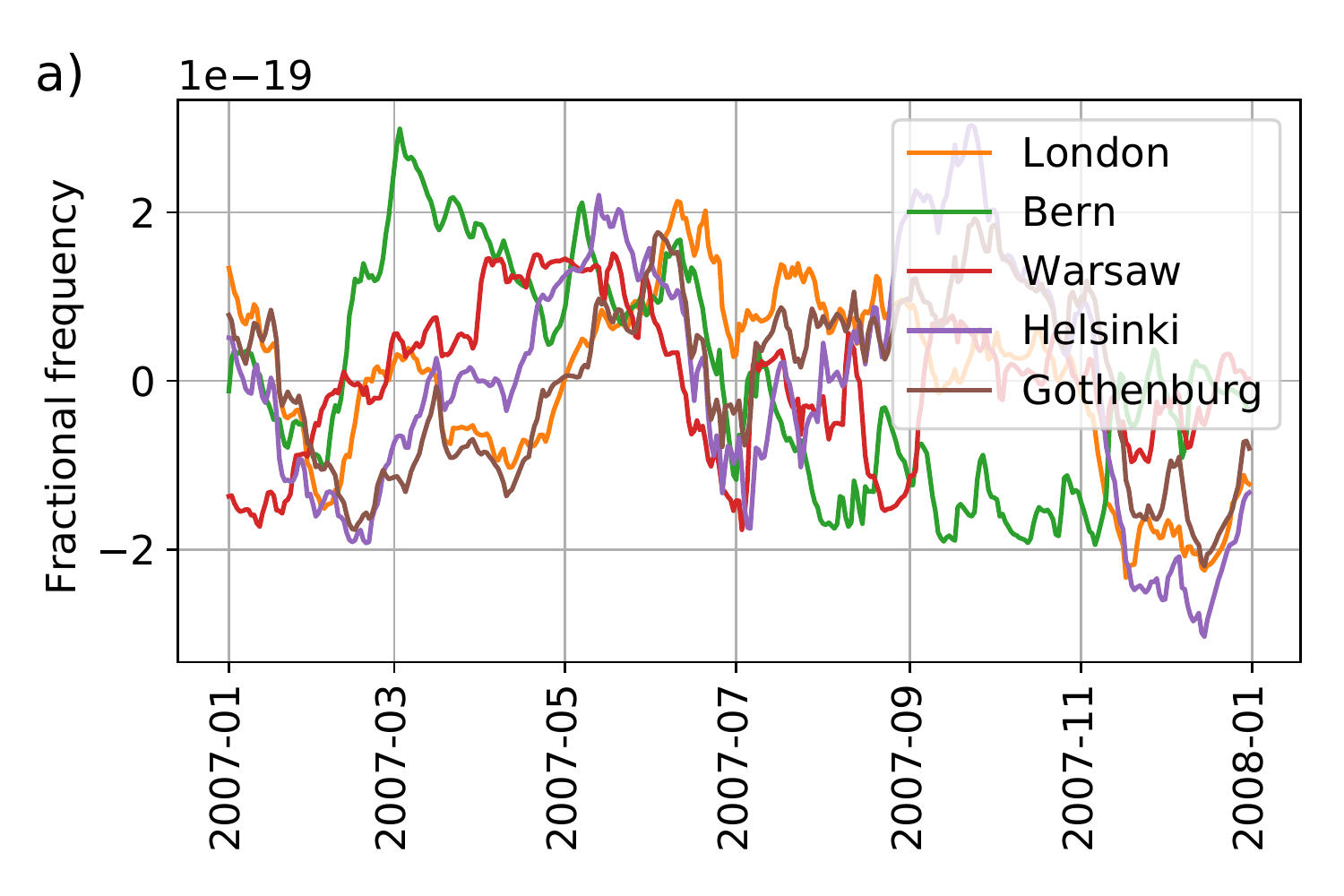}
\end{subfigure}
\begin{subfigure}[c]{0.49\textwidth}
\includegraphics[width=\textwidth]{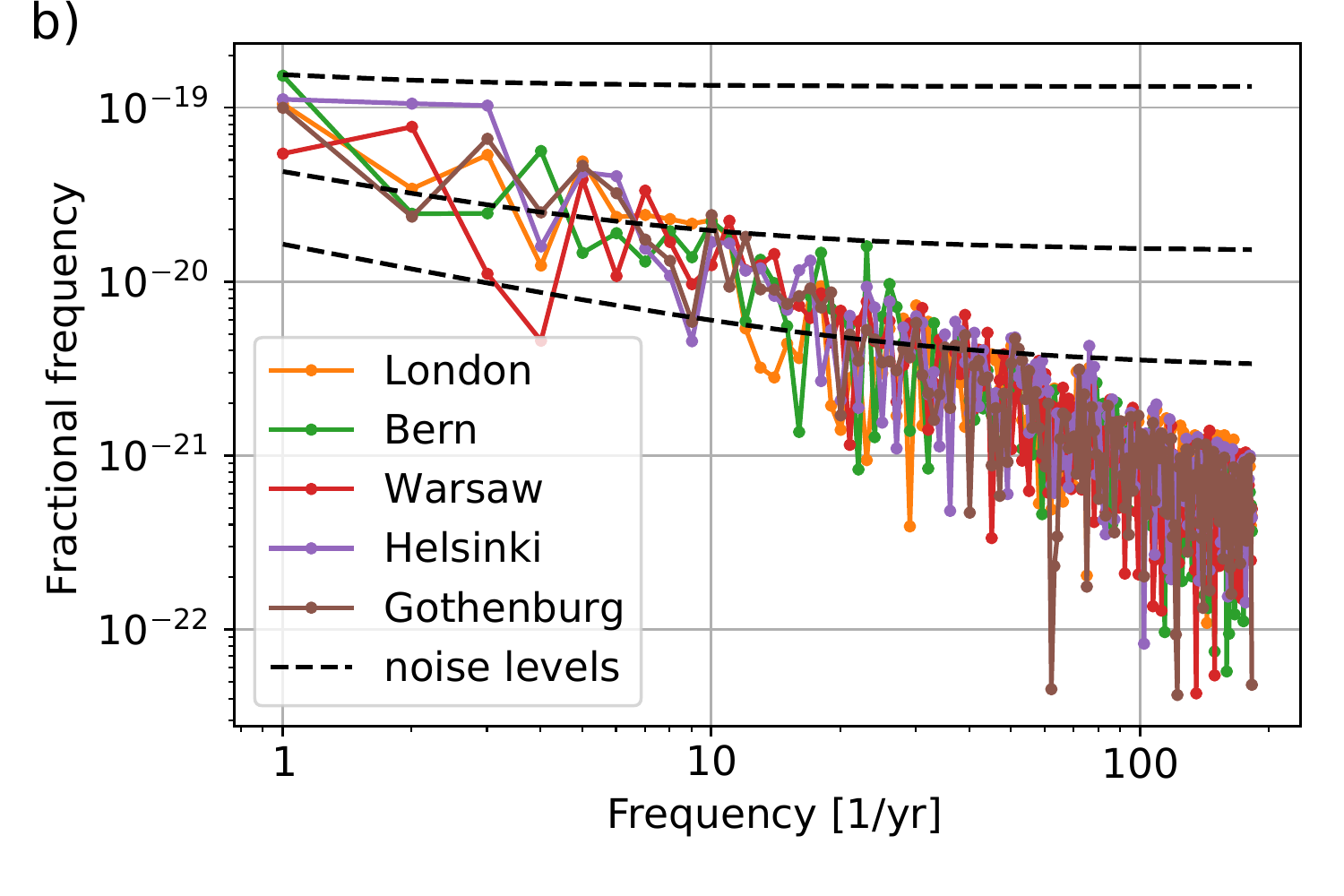}
\end{subfigure}
\caption{a) Fractional frequency variations induced by hydrological mass redistribution relative to \textbf{Braunschweig}. b) Corresponding amplitude spectrum including the three noise levels for clock comparisons.}
\label{fig:H_relative}
\end{figure}
In order to review this statement drawn from a single link, we show time series and amplitude spectra of several stations compared to Braunschweig in Fig. \ref{fig:H_relative}. Although the time series differ substantially -- now that a large part of the common annual signal is removed within the differencing approach -- their spectra all resemble closely the one predicted for the simulated link Bern vs. Braunschweig analysed before.
\par
Although individual peaks are almost all below $10^{-19}$, one will notice that after integration the signal level amounts to about $5\times 10^{-19}$. For comparison, \citet{voigt_time-variable_2016} report ranges of $10^{-18}$ over Europe in 2003-2004 (they used spherical harmonic coefficients up to degree and order 100 based on the GLDAS model), which is in alignment with what we show in Figure \ref{fig:H}. For differences at clock locations over continental scales they report $10^{-18}$ as well, which is twice of what we find (compare Fig. \ref{fig:H_relative}). This may be related to a slightly different simulation setup; our study assumed a network which does not cover the entire continent, and due to the differencing approach this will inevitably lead to some limitation in the maximum signal level. Generally, one can conclude from the fractional frequency simulations that short term hydrology , i.e. water storage changes at time scales below one month, can be neglected even for clock comparisons in scenario 2. In other words, given noise levels under scenario 2, they would not be detectable with such uncertainties of the clocks and GNSS. This conclusion would have to be revisited if one would bring a clock closer to a location in the Alps or the Scandinavian mountains though.

\subsection{Atmospheric mass changes}
\begin{figure}
\includegraphics[width=\textwidth]{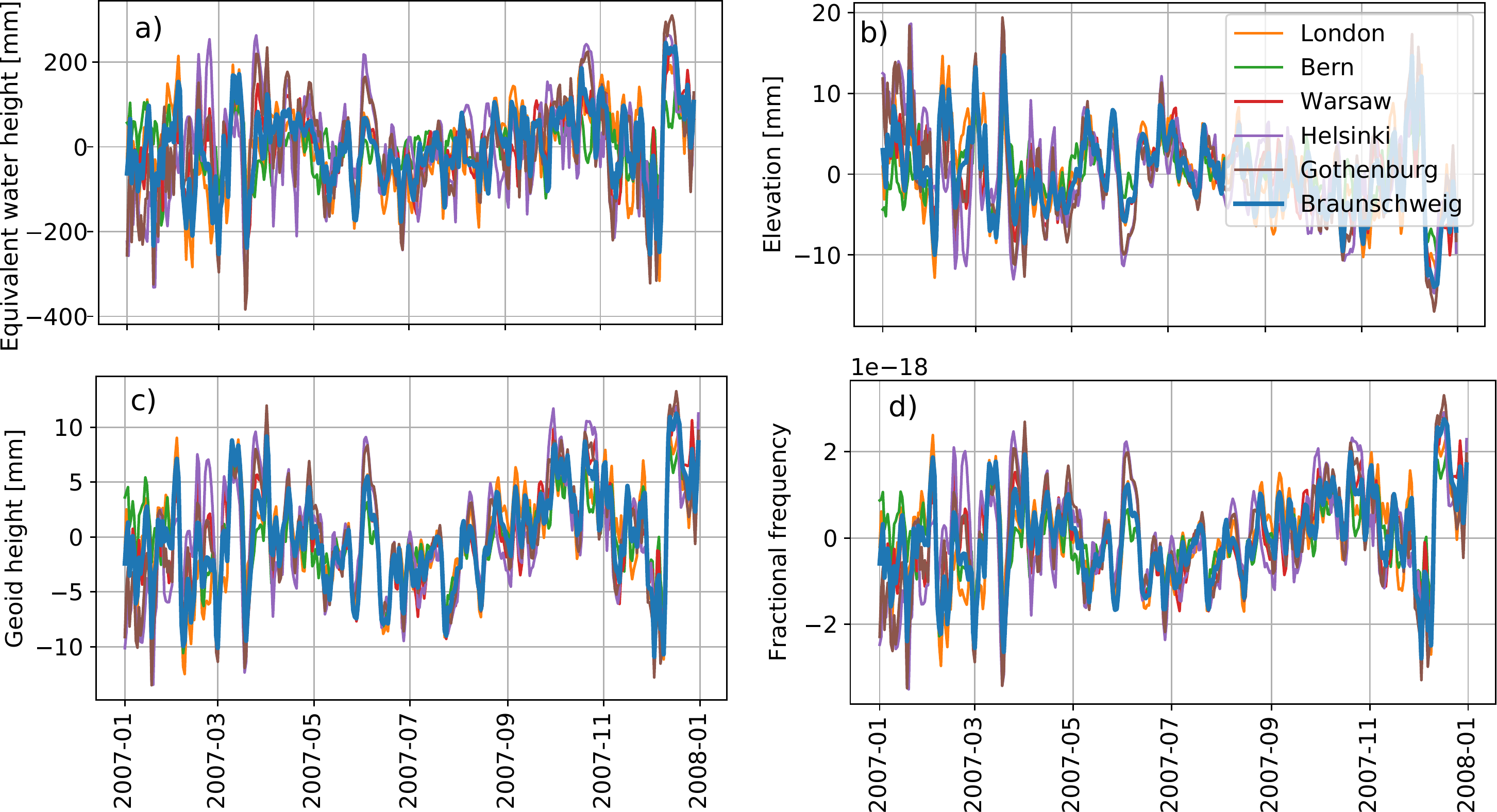}
\caption{a) Daily atmospheric mass variations at several stations in 2007 and resulting b) elevation, c) geoid height, and d) fractional frequency variations.}
\label{fig:A}
\end{figure}
In this section we will analyse the effect of atmospheric mass variability, which is both due to the redistribution of dry air masses and the corresponding pressure changes, and due to much faster water vapour variability. 
As the atmospheric mass variability is known to contribute much more to observed signals as compared to hydrological variability, Fig. \ref{fig:A} reveals a much larger amplitude of EWH variation, with $0.7$ m about three to four times as high as the EWH amplitude induced by hydrological mass changes. This transfers to vertical displacement, geoid height, and fractional frequency as well. For our simulated network, we observe that maximum fractional frequency amplitudes lie in the range of $6\times 10^{-18}$. Furthermore, there is no obvious annual signal observable, while the short term variations appear very high.
\par
\begin{figure}
\begin{subfigure}[c]{0.49\textwidth}
\includegraphics[width=\textwidth]{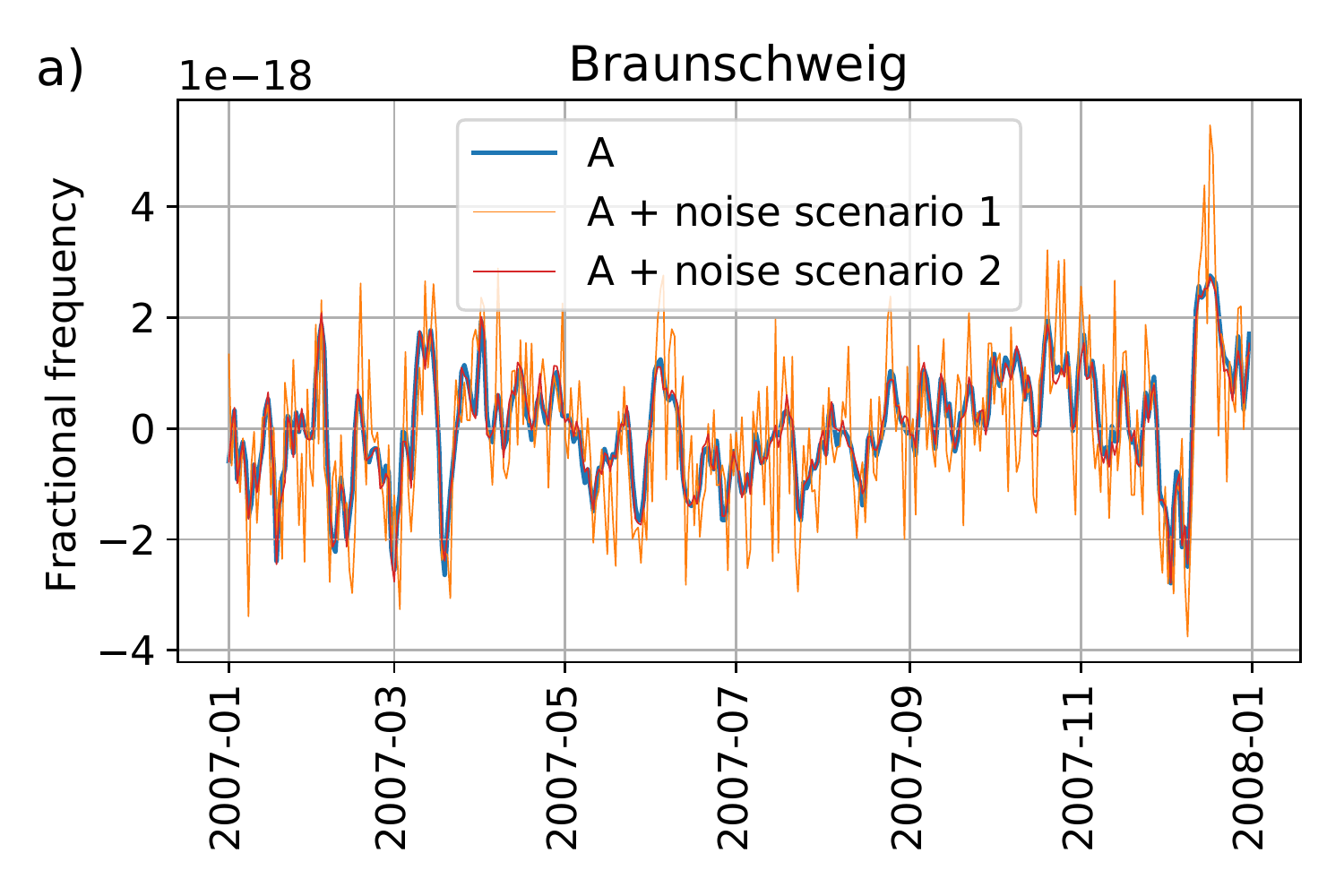}
\end{subfigure}
\begin{subfigure}[c]{0.49\textwidth}
\includegraphics[width=\textwidth]{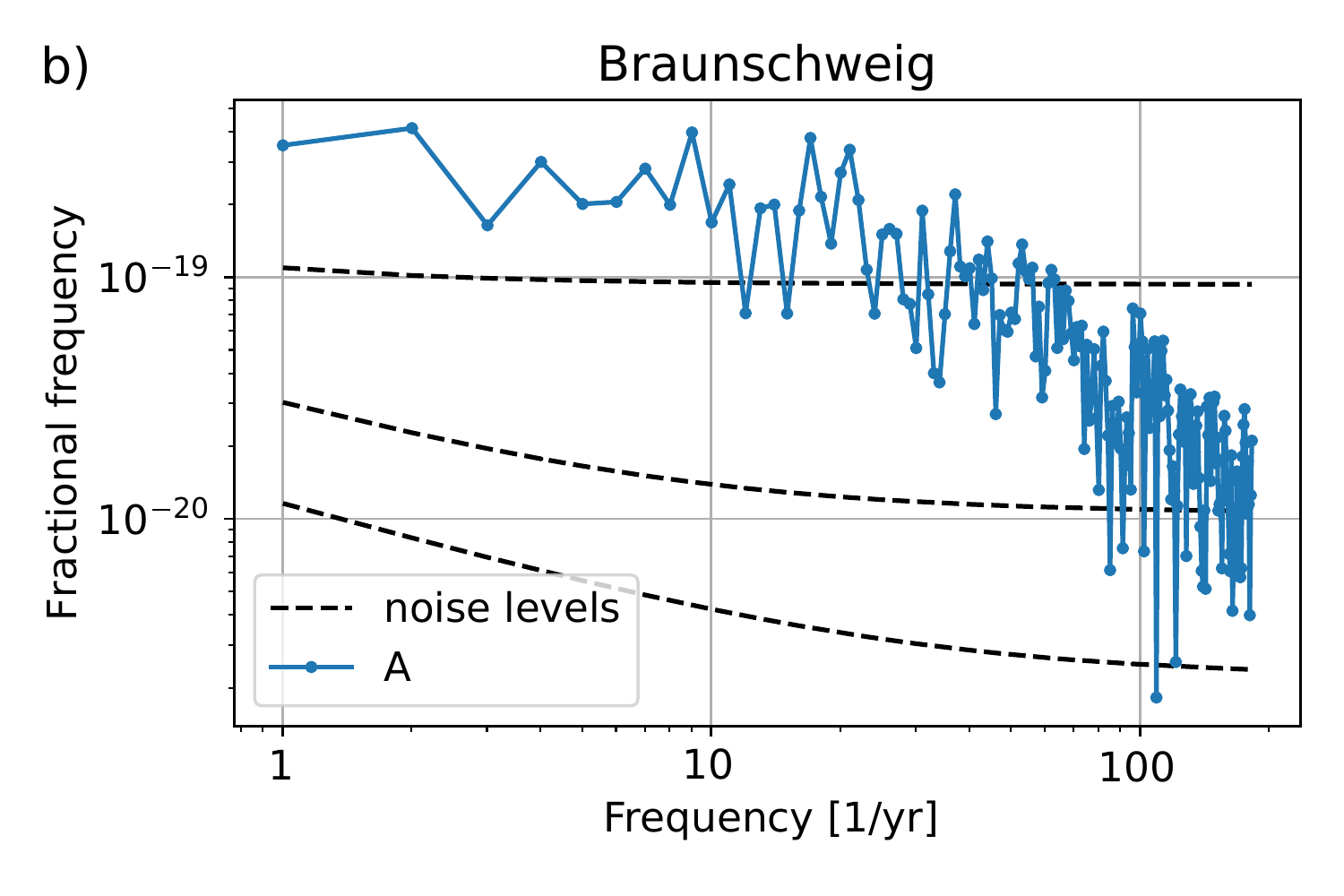}
\end{subfigure}
\caption{a) Time series of atmospheric mass induced fractional frequency variations in \textbf{Braunschweig} and b) the spectral domain of it.}
\label{fig:A_braunschweig}
\end{figure}
\begin{figure}
\begin{subfigure}[c]{0.49\textwidth}
\includegraphics[width=\textwidth]{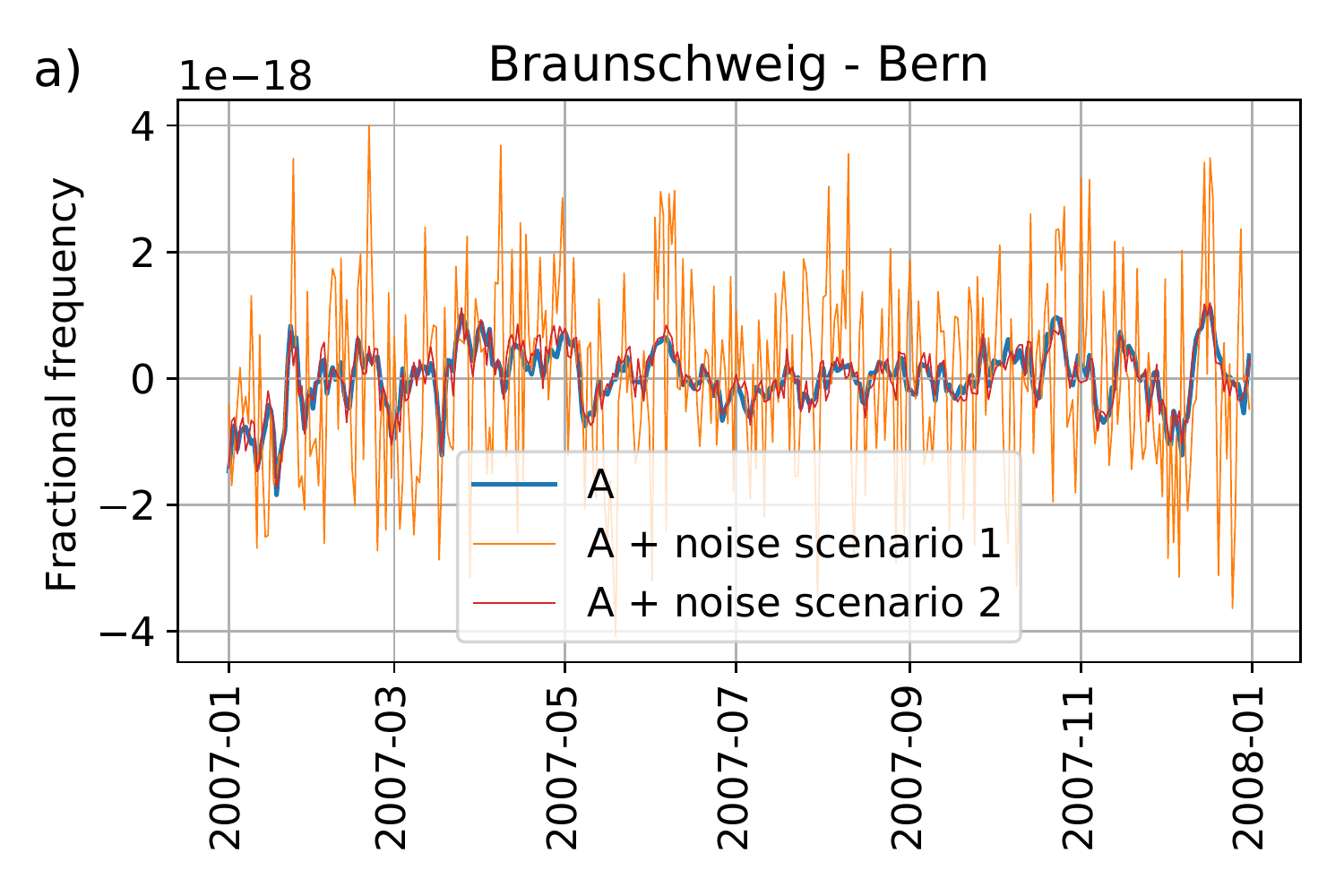}
\end{subfigure}
\begin{subfigure}[c]{0.49\textwidth}
\includegraphics[width=\textwidth]{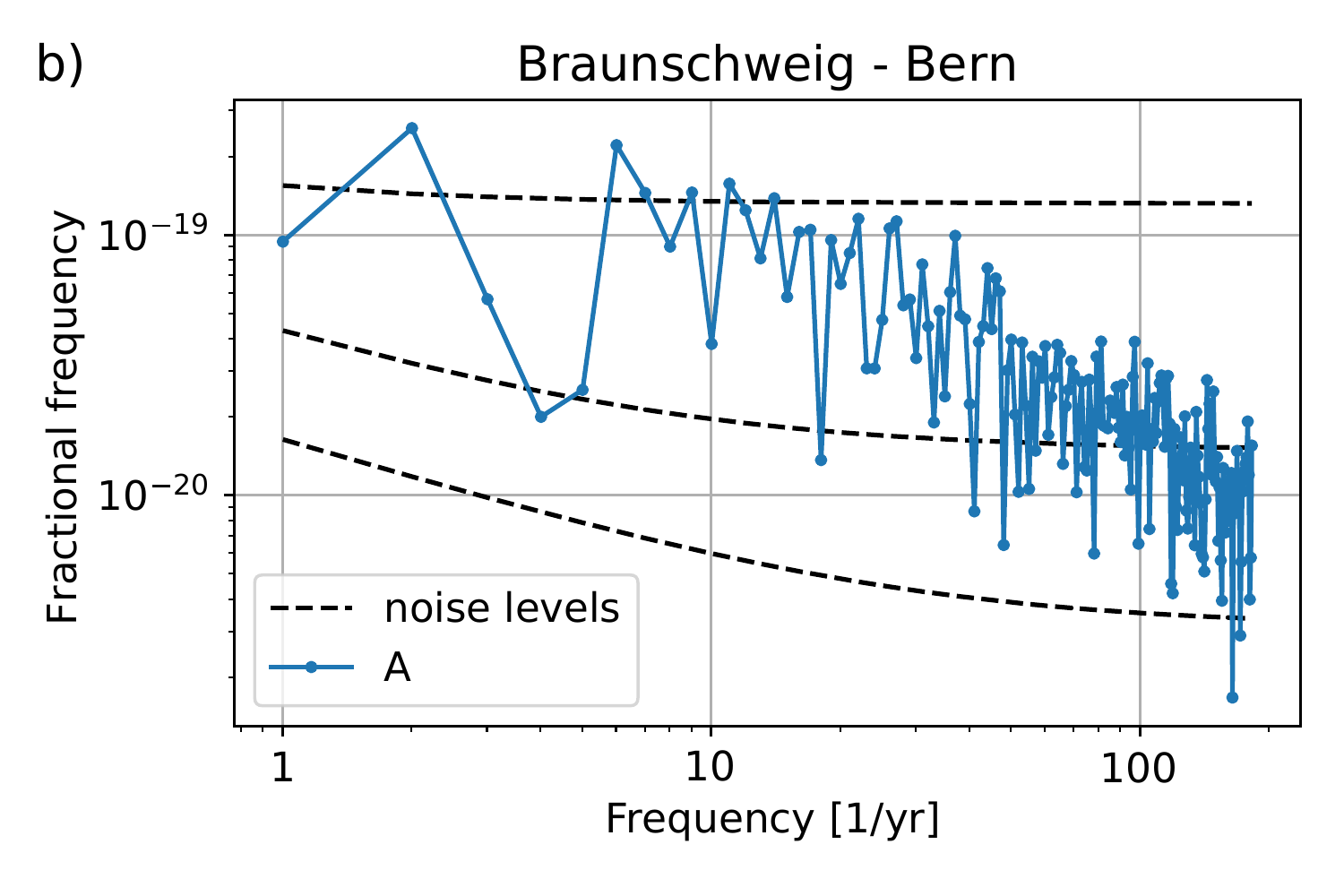}
\end{subfigure}
\caption{a) Comparison time series of atmospheric mass induced fractional frequency changes for the link \textbf{Bern vs. Braunschweig} and b) the spectral domain of it.}
\label{fig:A_braunschweig_bern}
\end{figure}
In Fig. \ref{fig:A_braunschweig} we show the simulated time series at the single clock location of PTB, Braunschweig again. The noise in scenario 1 is not as dominant as it appears compared to the hydrological signal, because the signal-to-noise ratio is higher. Also, up to a frequency of circa 30 cpa (cycles per annum) most of the signal spectrum appears well above the noise level for all three scenarios. When observing the simulated clock comparison times series and amplitude spectrum between Braunschweig and Bern (Fig. \ref{fig:A_braunschweig_bern}), one notices that in particular, the low frequency signal is reduced as compared to the single clock signal. Still, up to about 100 cpa the signal exceeds the noise floor of scenario 2.
\par
% \begin{itemize}
%     \item Fig. \ref{fig:A_braunschweig_bern}, now for the link instead of a single clock
%     \item signal is smaller, about half of the single clock signal
%     \item noise is larger by $\sqrt{2}$
%     \item thus, $\sigma$=1-e18 is again critical, only the peaks at 2 yr$^{-1}$ and 6 yr$^{-1}$ are well above this noise level
% \end{itemize}
% \begin{figure}[ht]
% \includegraphics[width=\textwidth]{figures/A_partofnetwork_ff_both.pdf}
% \caption{a) Atmospheric mass induced fractional frequency variations relative to \textbf{Braunschweig}. b) Corresponding spectral domain including the three noise levels for clock comparisons.}
% \label{fig:A_relative}
% \end{figure}
\begin{figure}
\begin{subfigure}[c]{0.49\textwidth}
\includegraphics[width=\textwidth]{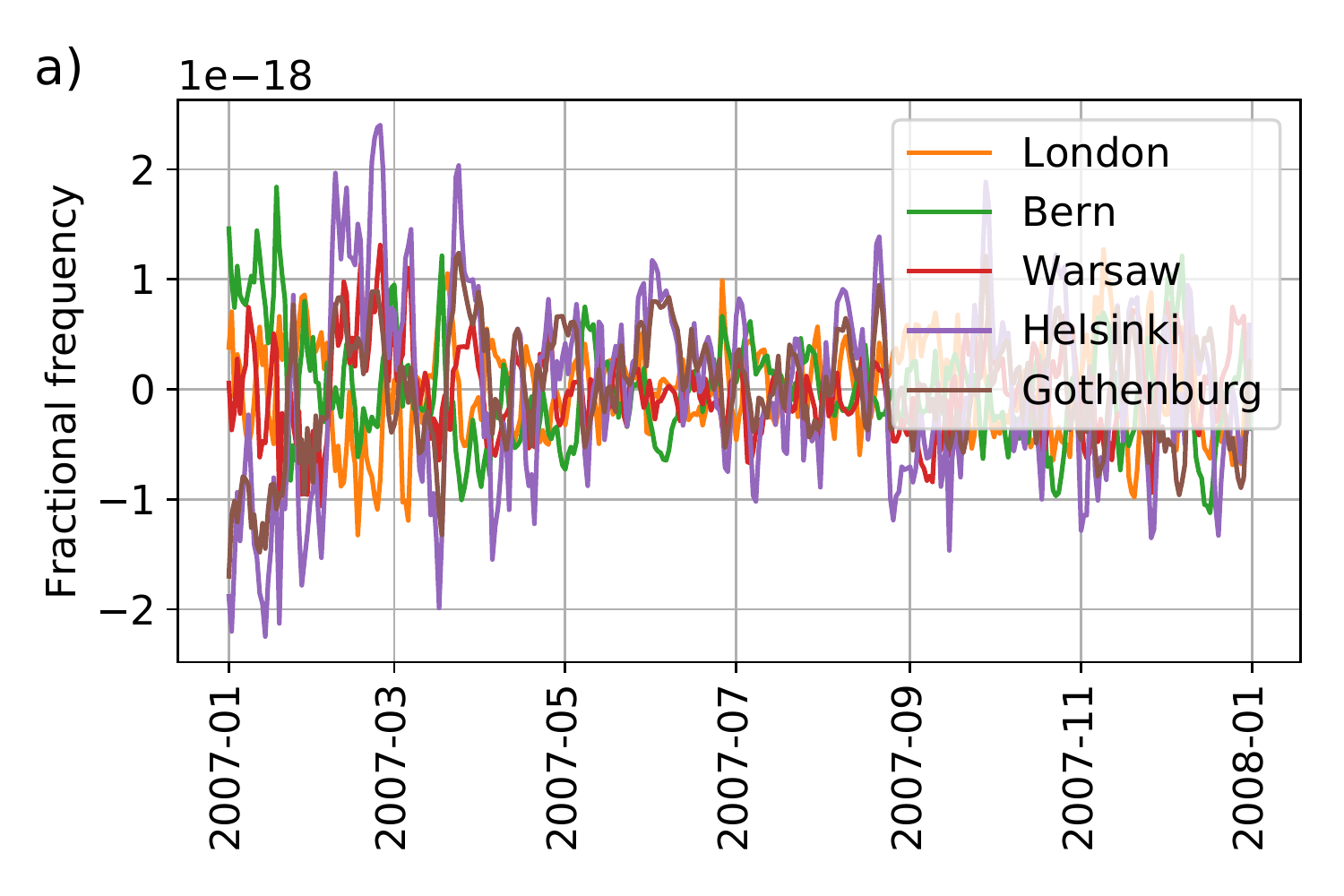}
\end{subfigure}
\begin{subfigure}[c]{0.49\textwidth}
\includegraphics[width=\textwidth]{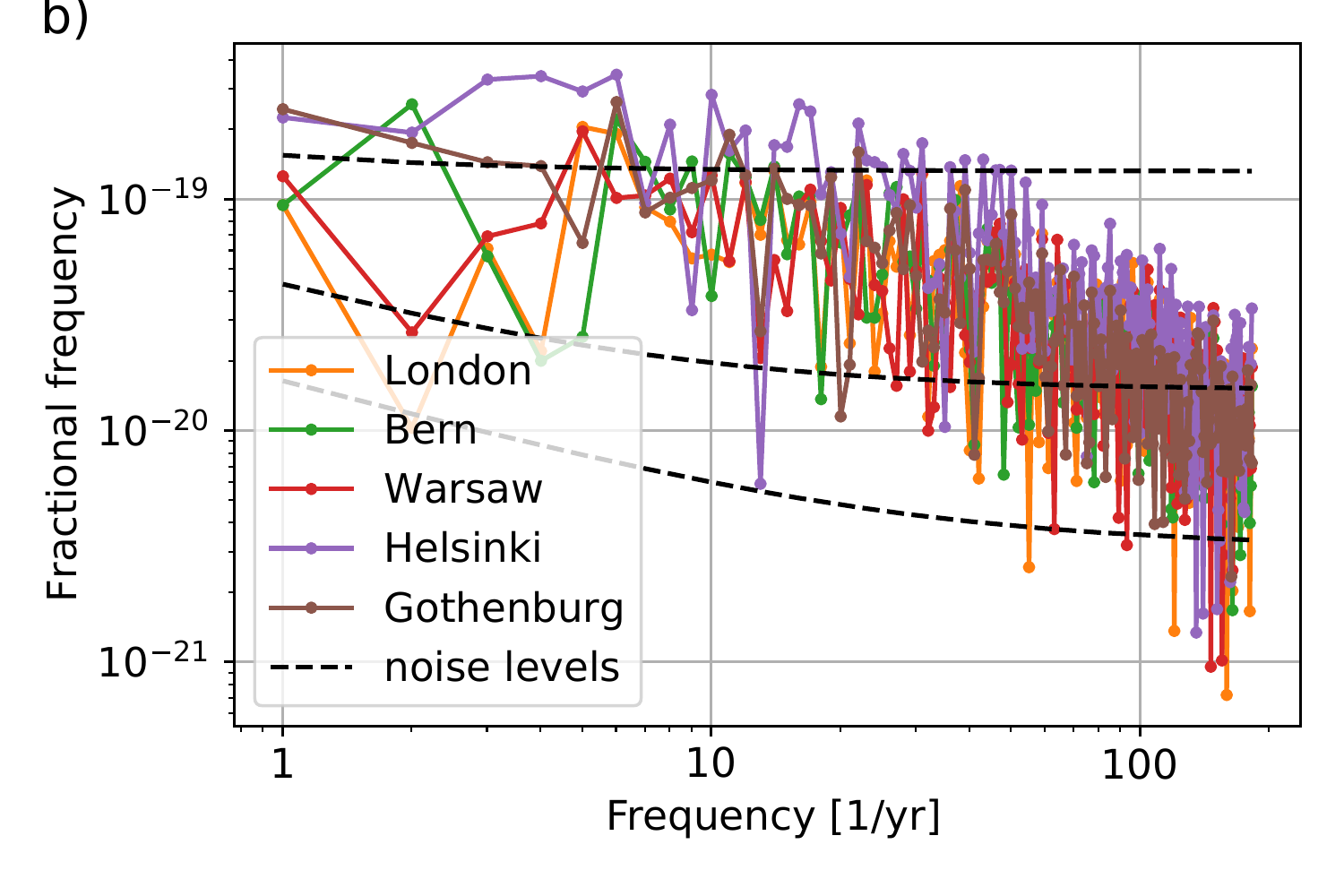}
\end{subfigure}
\caption{a) Atmospheric mass induced fractional frequency variations relative to \textbf{Braunschweig}. b) Corresponding spectral domain including the three noise levels for clock comparisons.}
\label{fig:A_relative}
\end{figure}
When comparing simulated fractional frequency differences for the link Bern vs. Braunschweig to differences for the other links (Fig. \ref{fig:A_relative}), the connection between MIKES (Helsinki) and Braunschweig stands out. With 1250km distance, Helsinki is the location furthest away from Braunschweig. This highlights the influence of distance between the clocks, something that we did not observe to this extent for the hydrological effect. For a clock comparison along the link Helsinki vs. Braunschweig, one would find fractional frequency variations of up to $5\times 10^{-18}$, thus almost as large as the single clock variations. \citet{voigt_time-variable_2016} suggest geoid height and vertical displacement values during a storm surge in the North Sea which would add up to $2.5\times 10^{-18}$ relative to a mean value. This is similar to what we find ($3\times 10^{-18}$ deviation from the mean, e.g. during a deep depression leading to cyclone Kyrill in mid of January 2007). 
However, we mention in passing that, unlike noted in \cite{voigt_time-variable_2016}, the direction of the geoid change does not depend on whether the mass change is happening below the clock, i.e. groundwater and ocean mass change, or above the clock, i.e. glacial and atmospheric mass change. 
This can conceptualized through imagining a 'mass disk' below, at, or above the Earth's surface: This mask disk leads to a jump in the gravity acceleration, with positive sign above, and negative sign below itself. Since the potential is the integral over the gravity acceleration along the vertical \citep{hofmann-wellenhof_physical_2005}, it only leads to a cusp for the potential, with positive sign everywhere and maximum in itself.
Interestingly, the clock comparisons over the network distances do not show much smaller variations than the single clock time series. This clearly points out the importance of atmospheric mass correction data for European fibre link optical clock comparisons at the spatial and temporal scales that we consider here, and that would be relevant for validating a future gravity mission.

\subsection{Glacier mass variability}
\begin{figure}
\begin{subfigure}[c]{0.49\textwidth}
\includegraphics[width=\textwidth]{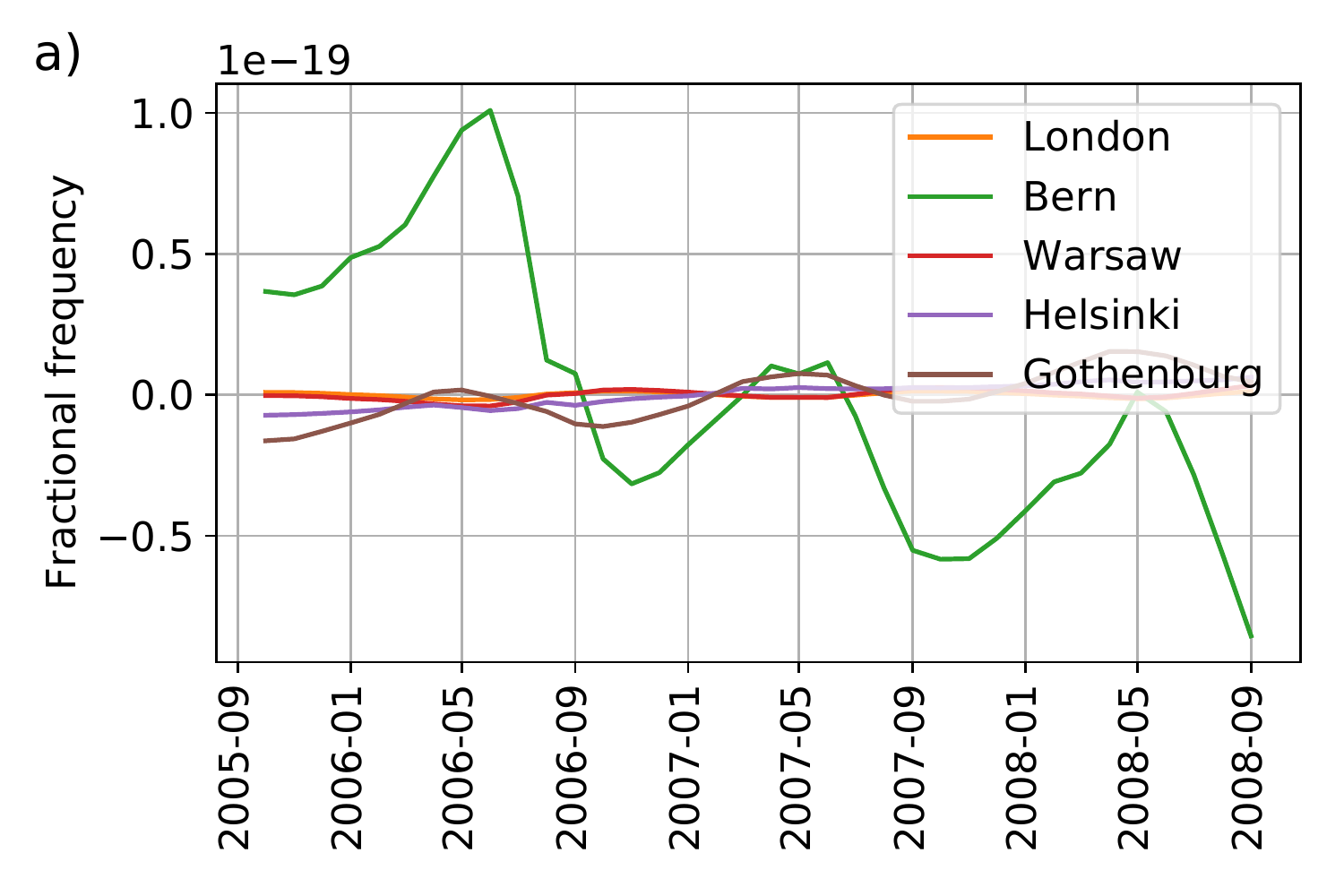}
\end{subfigure}
\begin{subfigure}[c]{0.49\textwidth}
\includegraphics[width=\textwidth]{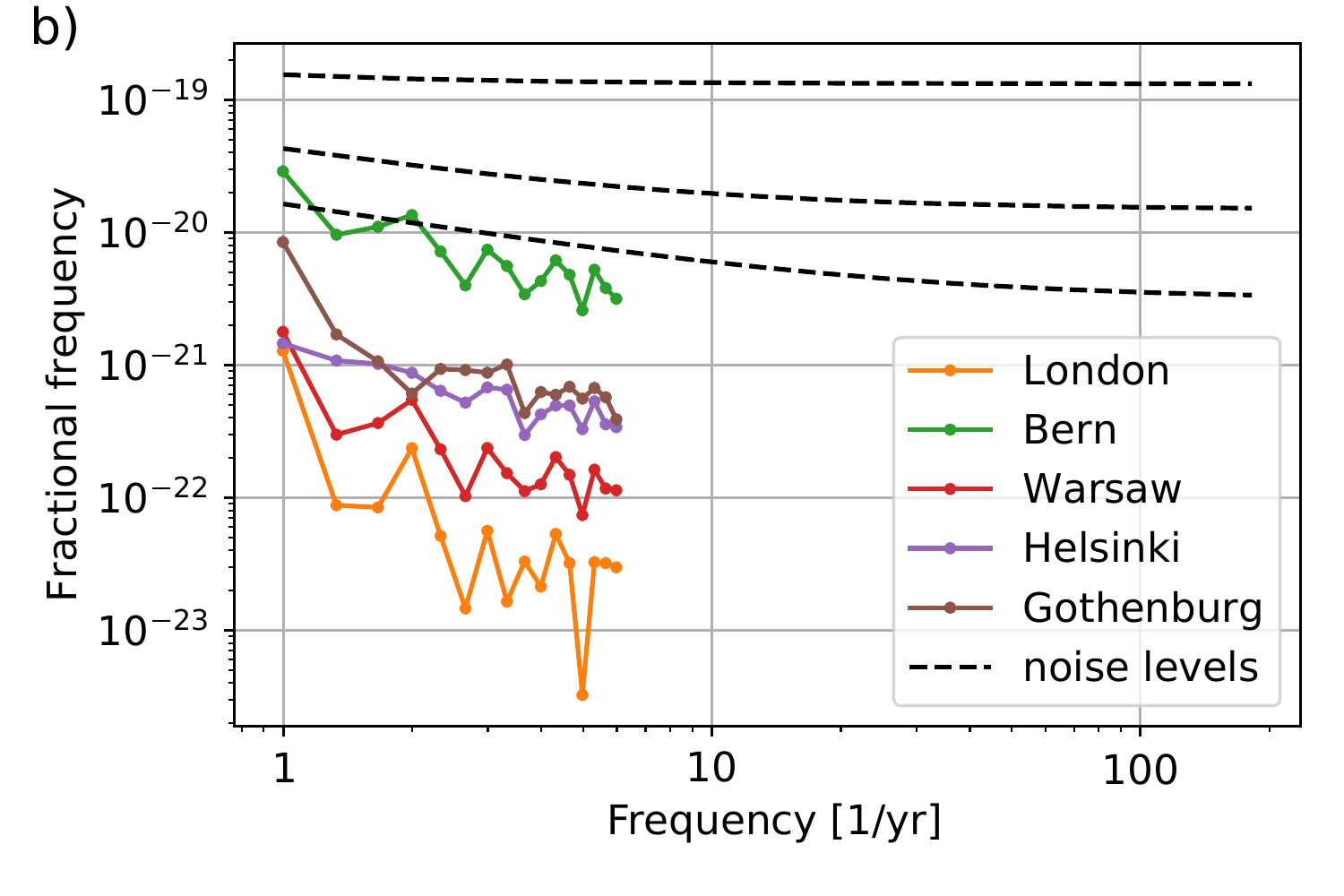}
\end{subfigure}
\caption{a) Glacier induced fractional frequency variations relative to \textbf{Braunschweig}. b) Corresponding amplitude spectrum including the three noise levels for clock comparisons.}
\label{fig:I_relative}
\end{figure}
Glacier models are typically run at monthly resolution, due to the slow response behaviour of the glaciers. As a result, we analyse monthly glacier mass changes here, dating from October 2005 to September 2008. Although the time series is very short for an analysis in the spectral domain, some results are obvious: In Figure \ref{fig:I_relative}a) the simulated clock link from Braunschweig to Bern displays the highest fractional frequency variability with a distinct annual amplitude up to $1.3 \times 10^{-19}$. Gothenburg shows an annual signal as well but to a much smaller extent. That is observable in the spectral domain as well, where Bern is the only location with single amplitudes exceeding the noise level of scenario 3. In conclusion, as long as the clock locations are not closer to glaciated areas than Bern is, the time variable signal of the glaciers will play only a minor role compared to hydrology and atmosphere induced mass redistributions. This will be true in the case of daily variations as well, as glacier changes are dominated by seasonal signal. What could become a factor, however, are long term trends, as glaciers undergo strong melting during global warming.
Moreover, the total mass imbalance for European glaciers is estimated to only about $2$ Gt yr$^{-1}$, compared to 73 Gt yr$^{-1}$ for Alaskan glaciers \citep{zemp_global_2019} or to the mass imbalance of the Greenland ice sheet, quantified to $-$272 Gt yr$^{-1}$ by the \citet{wcrp_global_sea_level_budget_group_global_2018}.
\section{Discussion}
Time-differencing clock comparisons measure differences in geopotential change with differential height changes superimposed. At the time being, using colocated, high-precision GNSS measurements appear the best choice for separating the geopotential. As we have shown, the GNSS uncertainty will soon become a limiting factor for clock comparisons on the ground. This requires additional discussions; we will focus on two potential remedies here:
\par
A straightforward option could be instead of relying on GNSS to co-estimate the elastic loading jointly with the geoid change, as it has been proposed for analysing tide gauge data (e.g. \citealp{mitrovica_quantifying_2018}). However this requires that other effects can be either corrected for (e.g. relying on models for glacial isostatic adjustment), or are small when compared to the signal of interest. While for validating large-scale mass redistribution at least over Europe this will hardly be the case, this idea could be explored for clocks that are brought deliberately close to large mass redistributions such as e.g. glaciers. Finally, since several subsidence processes evolve nearly linear in time one could simply rely on de-trending clock comparison time series; again the downside would be that most of the climate signal would be lost, unfortunately.
\par
\begin{figure}
\begin{subfigure}[c]{0.33\textwidth}
\includegraphics[width=\textwidth]{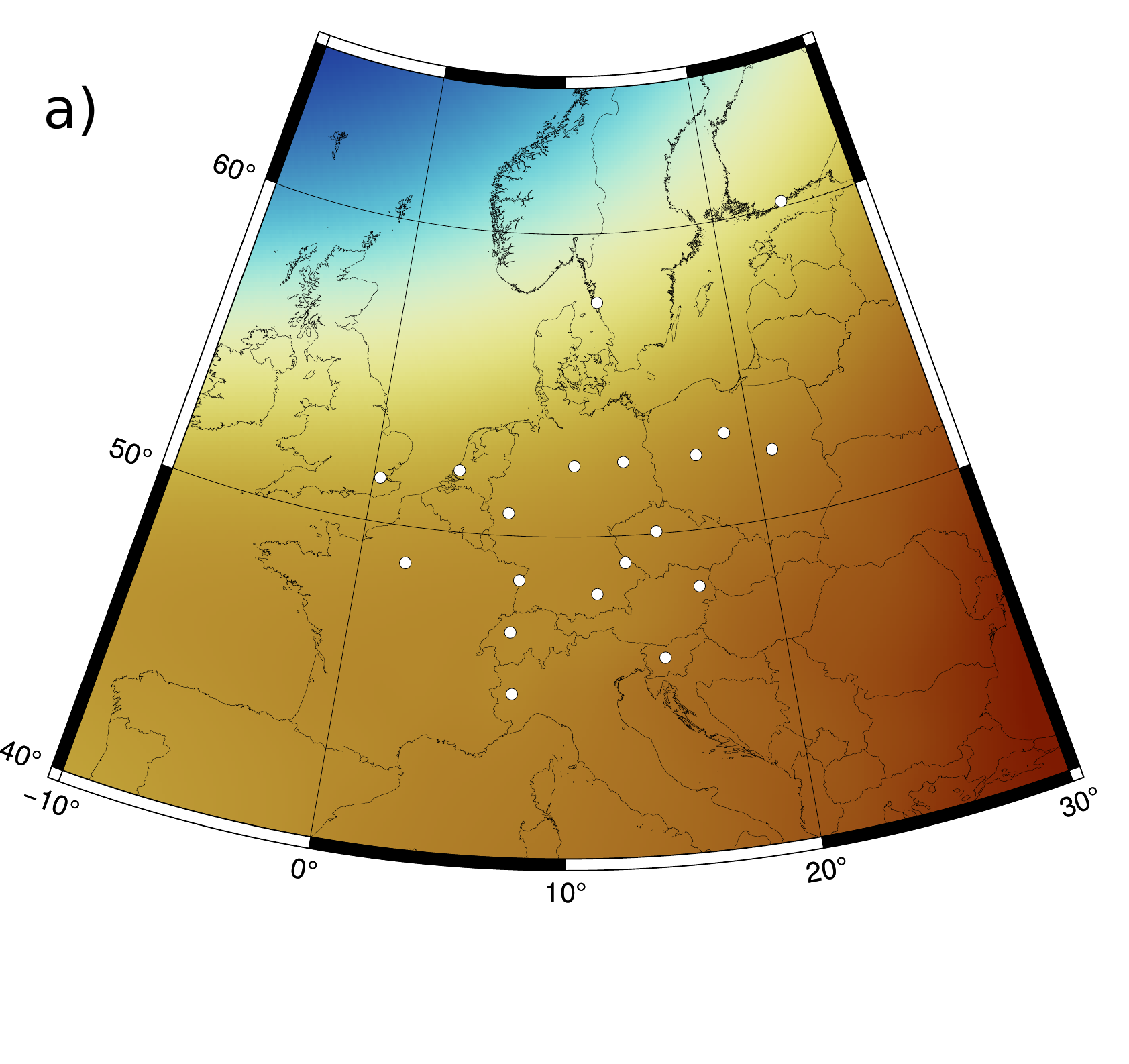}
\end{subfigure}
\begin{subfigure}[c]{0.33\textwidth}
\includegraphics[width=\textwidth]{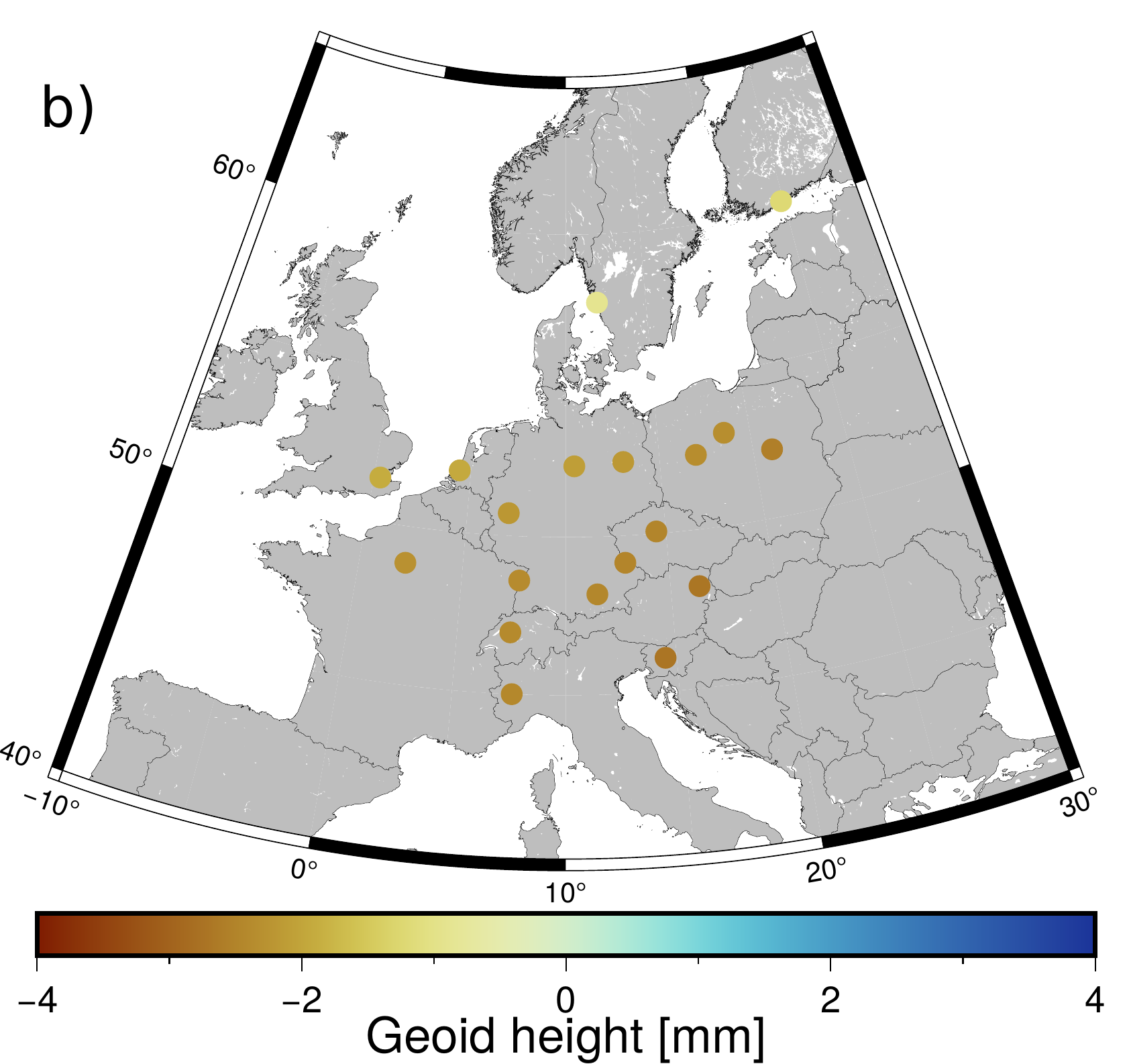}
\end{subfigure}
\begin{subfigure}[c]{0.33\textwidth}
\includegraphics[width=\textwidth]{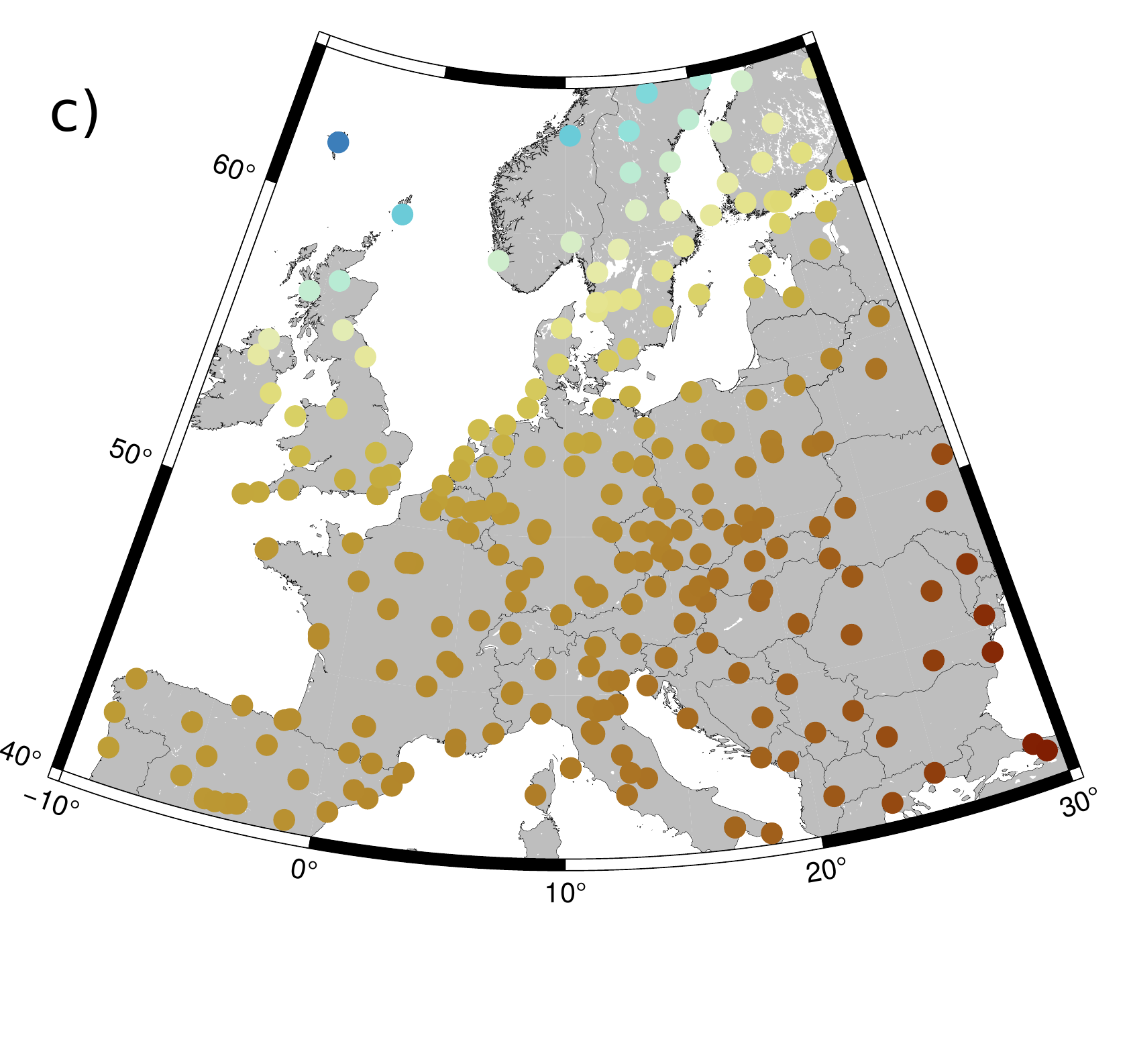}
\end{subfigure}
\caption{Visualization of geoid height anomalies in June 2007 with respect to the 2007 mean. a) visualizes the full signal, b) shows point-wise anomalies at the clock locations in our assumed network from Fig. \ref{fig:network}, and in c) the anomalies are shown at the locations of all EUREF Permanent Network (EPN) GNSS stations.}
\label{fig:A_euref}
\end{figure}

Assuming that optical clocks would reach an uncertainty limit corresponding to scenario 2, i.e. $10^{-19}$, within a few years from now, while at the same time becoming more affordable and easier to operate, a denser and more widespread network could be pictured (see Fig. \ref{fig:A_euref}). Increasing the number of clocks instead of their precision and accuracy might be a more desirable objective, mainly for the following reasons.
First, when GNSS becomes a limiting factor for clock comparisons, it will just not be beneficial to use better clocks. Moreover, while GRACE/-FO and possibly future satellite gravity missions will inevitably be limited in resolving a high spatial resolution (e.g. \citealp{pail_observing_2015, flechtner_what_2016, rodell_emerging_2018}), a sufficiently dense clock network could fill this gap.
In order to demonstrate the potential benefit of this strategy, Fig. \ref{fig:A_euref} visualizes geoid height changes caused by the atmospheric mass redistribution in June 2007 with respect to the 2007 mean. While in panel a) one can see the full signal, in b) the geoid-height anomalies are only shown point-wise, at the clock locations of our simulated network (compare Fig. \ref{fig:network}). In Fig. \ref{fig:A_euref}c) we assume a denser network; geoid-height changes are here shown at all EUREF Permanent Network (EPN) GNSS stations \citep{bruyninx_gnss_2019}. The shorter distances between the almost 300 station locations would allow for gravity potential estimation with a spatial resolution even beyond what multi-pair satellite gravity missions are expected to provide \citep{elsaka_comparing_2014}.
\par
\begin{figure}
\begin{subfigure}[c]{0.49\textwidth}
\includegraphics[width=\textwidth]{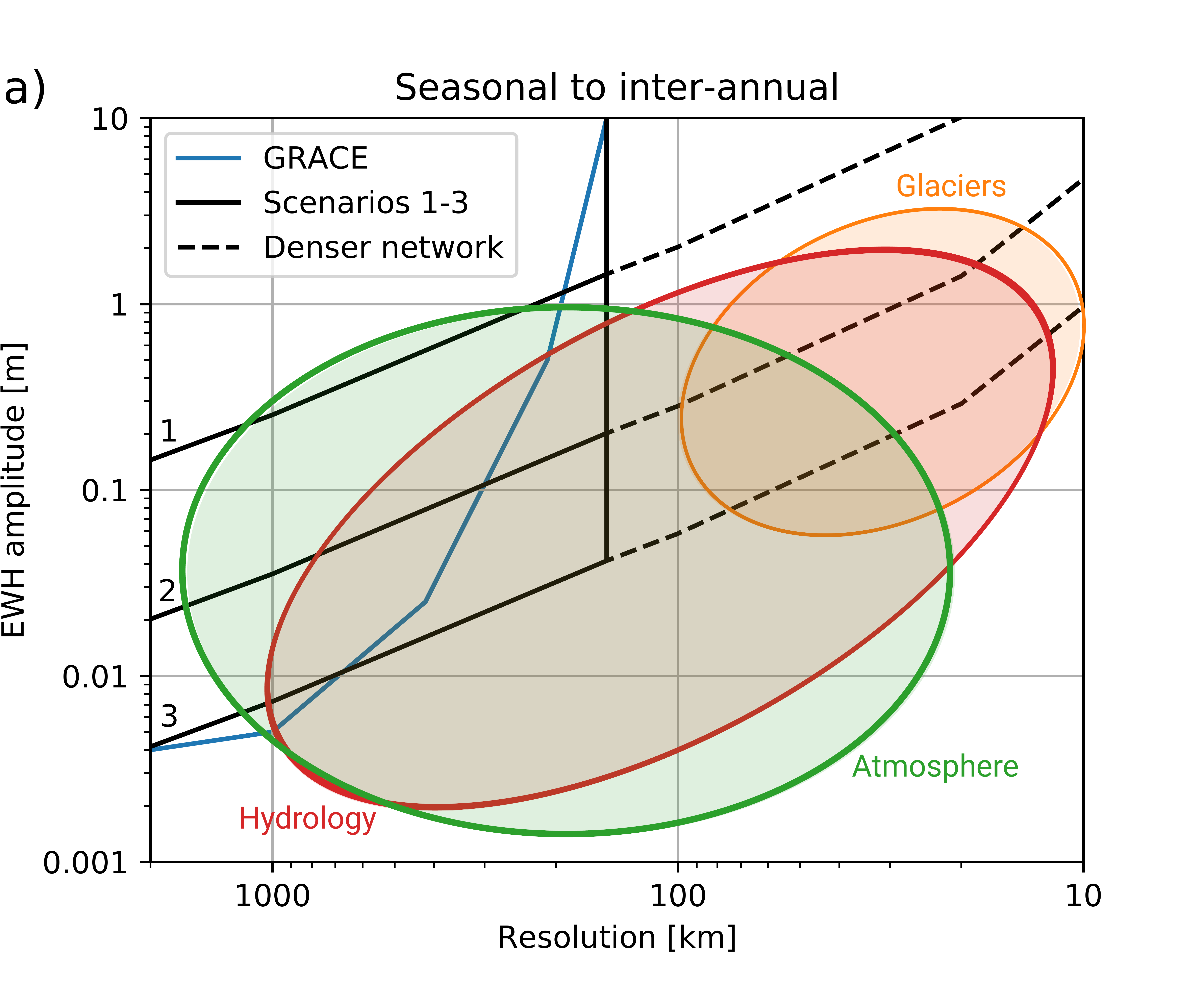}
\end{subfigure}
\begin{subfigure}[c]{0.49\textwidth}
\includegraphics[width=\textwidth]{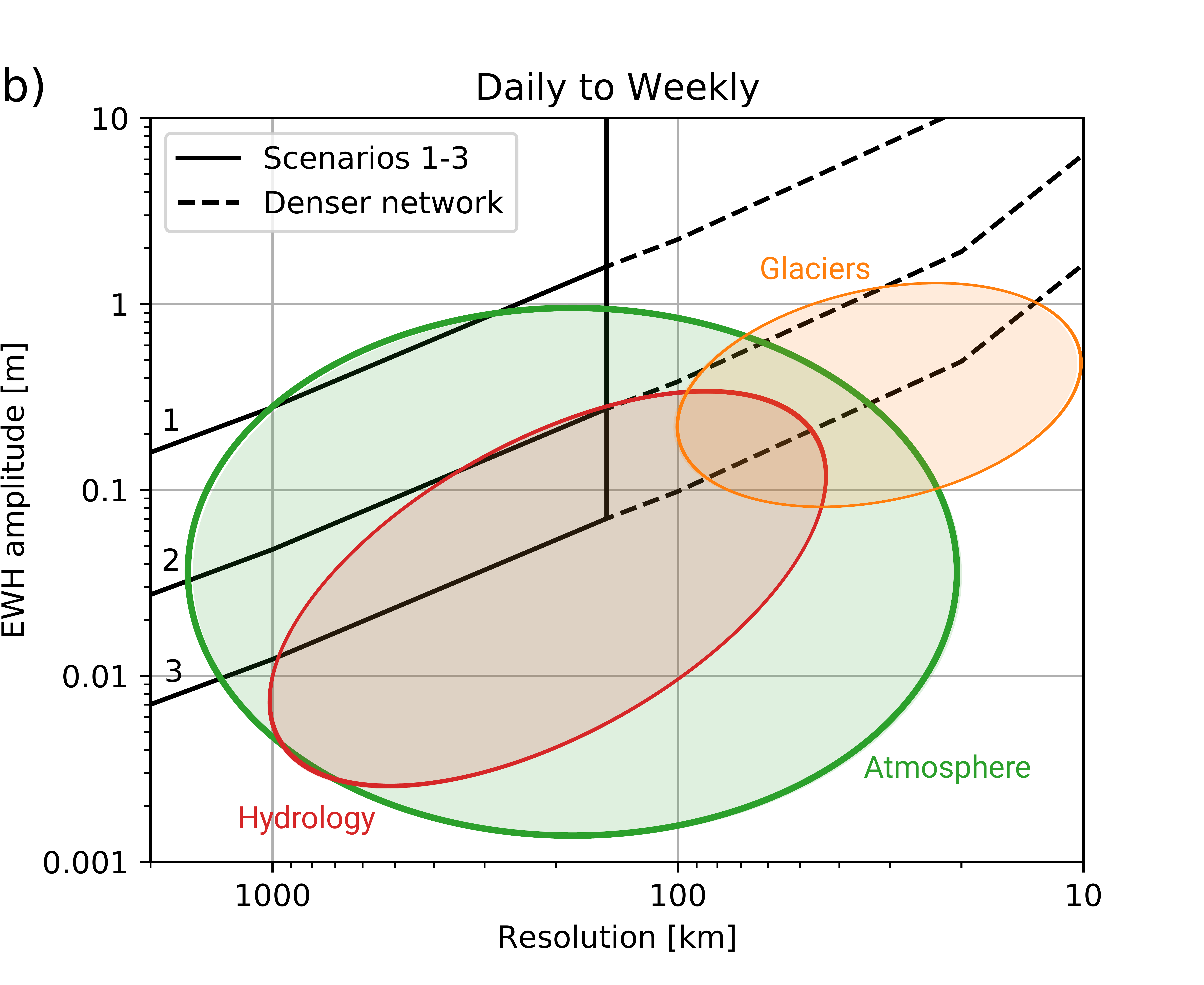}
\end{subfigure}
\caption{Magnitude versus spatial scale of the here discussed mass redistribution signals for two different temporal scales.}
\label{fig:bubble}
\end{figure}
We confront the expected spatial scale and magnitude of time-variable gravity signals against the uncertainty of clock network comparisons with associated GNSS height control within the three scenarios that we considered (see Table \ref{tab:scenarios}) in Fig. \ref{fig:bubble}, for two different temporal scales similar to \citet{pail_observing_2015}. The GRACE uncertainty (a) here is taken from \citet{pail_observing_2015}. The network resolution limits are derived here from Fig. \ref{fig:network}, where the longest distance (Helsinki - Torino) is about 2000 km. We have assumed four links (Warsaw - Torun, Torun - Posen, Munich - Wettzell, Braunschweig - Potsdam) slightly below 200 km distance, where we set the lower bound for the clock network resolution in the plot. Dashed lines would visualize the detectability of mass changes by a densified network as suggested in this Section.
We assume that the spatial resolution has no effect on the uncertainty of geoid height changes measured by the clock network, since we assume the length of the fibre links to be irrelevant. 
We note that in Fig. \ref{fig:bubble} the network's uncertainty in detecting EWH changes increases with increasing resolution, however the reason for that is merely that mass redistributions at smaller scales have smaller influence on the geoid (compare Eq. 3-5).
GRACE, however, exhibits a steeper increase of uncertainty with increasing resolution, mainly due to a higher signal-to-noise ratio resulting from its altitude.
Thus, it becomes clear that an optical clock network would observe contributions of hydrologic and atmospheric mass changes which are likely not visible in current GRACE/-FO data, and which would possibly also not be visible in data obtained from future gravity missions. At daily to weekly scales (Fig. \ref{fig:bubble}b), we do not show the GRACE curve as the standard release of GRACE products is of monthly resolution. This may change with future gravity missions though, at least for larger spatial scales and depending on their configuration and instrumentation.
On the other hand, Fig. \ref{fig:bubble}b) reveals that -- at least over Europe -- the densification of the clock network of scenario 1 would not be beneficial, because no loading signals on the smaller scales are strong enough to be detected.
\par
\begin{figure}
\includegraphics[width=\textwidth]{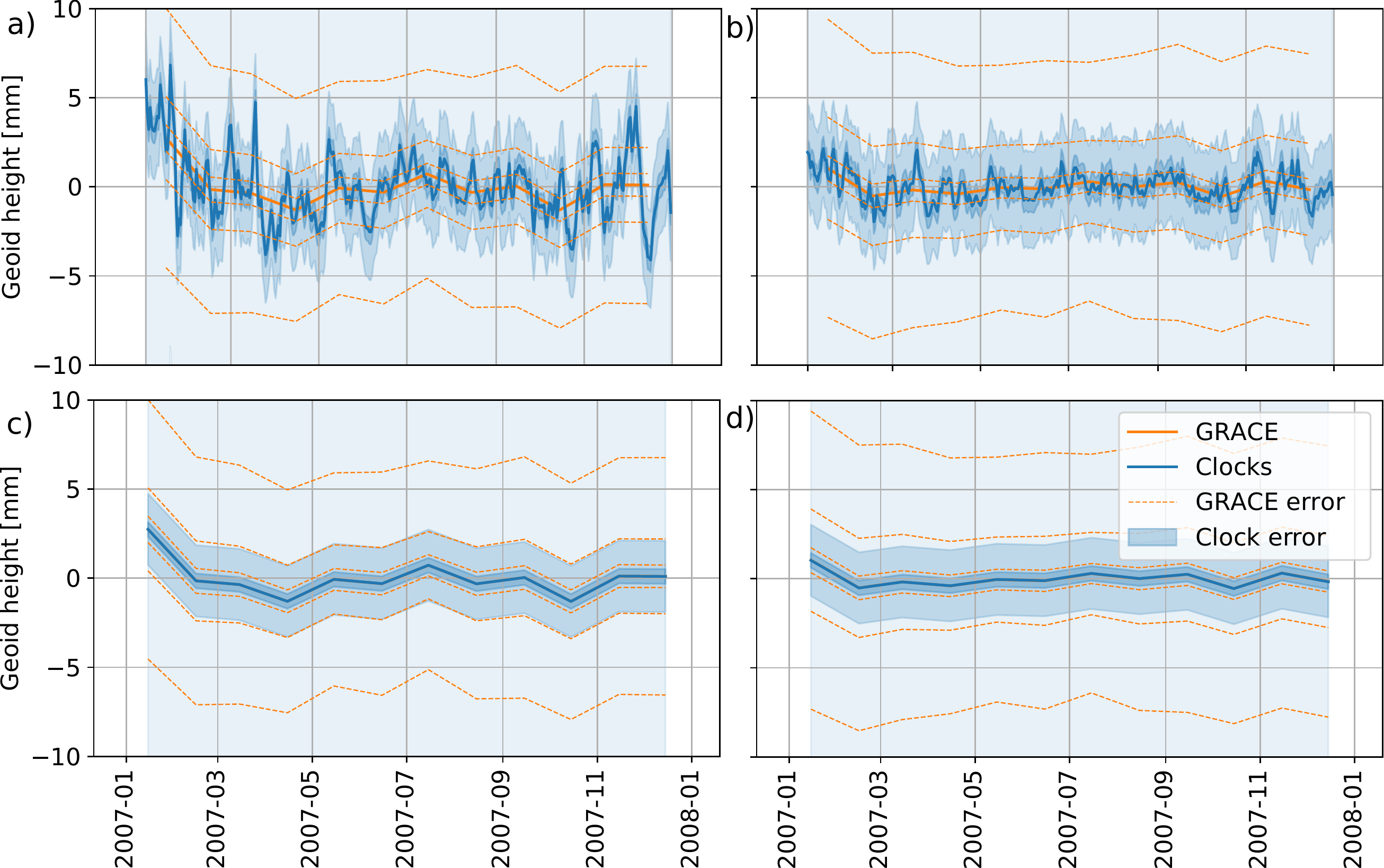}
\caption{Time series of atmosphere induced geoid height variations of \textbf{Bern vs. Braunschweig} in the first column and \textbf{Bonn vs. Braunschweig} in the second column. Comparison of the time series as would be seen from clock comparisons and simulated GRACE data. In the second row, the daily clock measurements are averaged to a monthly resolution. 1$\sigma$ error bounds are added, for the three different clock network error scenarios in transparent blue, and for the GRACE errors with dashed orange lines. The GRACE errors are ITSG-2018 formal errors, truncated at degrees 60 (smallest error bounds), 80, and 100 (largest error bounds), propagated to the geoid height difference between the two locations.}
\label{fig:A_clocks_vs_grace_ts}
\end{figure}
Fig. \ref{fig:A_clocks_vs_grace_ts} shows a comparison of the simulated clock links Bern - Braunschweig (left) and Bonn - Braunschweig (right) vs. simulated GRACE measurements, where the simulated signal is generated here only by the atmospheric variability.
The transparent blue areas display the error bounds of the clock comparison under the three scenarios, while scenario 1 leads to error bounds beyond the axis limits here. In contrast to that, we demonstrate GRACE error bounds resulting from the error propagation of ITSG-2018 formal errors. We have decided against applying a filter because it corrupts the comparibility of the GRACE errors and the point-wise clock errors. This is due to the fact that the GRACE errors refer to an area around a point instead of a point; the size of the area depends on the maximum spherical harmonic degree, while for point errors one would need infinitely many SHC's.
Instead we truncated the spherical harmonic coefficients at different degrees. Truncating the SHC's at degree 60 leads to error bounds at roughly the same extension as the error bounds from the clock scenario 3, i.e. with $10^{-20}$ clock uncertainty and $0.3$ mm GNSS uncertainty. This can be observed more distinctly when the clock comparison time series is averaged to a GRACE-like monthly resolution. If the maximum spherical harmonic degree of the GRACE formal errors is increased to 80, the error bounds increase as well and are in the same range as the error bounds from clock scenario 2, i.e. with $10^{-19}$ clock uncertainty and $0.7$ mm GNSS uncertainty.
We suggest that a time series comparison between a clock comparison and satellite gravimetry observations would have to be conducted applying different truncation degrees and possibly by correcting clock readings with local or regional model data. 
\par
It has been proposed that terrestrial water storage anomalies (TWSA), as an Essential Climate Variable (ECV), would be monitored from space at least with spatial resolution of 300 km at monthly timescale, and with a measurement accuracy of $10$-$20$ mm monthly and $10$ mm yr$^{-1}$ for trends. Assuming that this spatial resolution may be met by a clock network, this translates into an uncertainty of $0.1$ to $0.2$ mm necessary for the network's overall geoid height estimation, so slightly below our assumed scenario 3.
\par
In order to validate -- over Europe -- a future gravity mission that we assume is about as precise as GRACE, a clock network would require a comparison error not greater than $2$ mm geoid height, including link performance and GNSS control. This would work mostly for spatial scales of 200 to 400 km, while beyond that the GRACE precision exceeds the limits of the proposed clock network. Assuming that the future gravity mission is about five times as precise as compared to GRACE, the same clock network could be used for validation at 200 to 300 km spatial resolution of the gravity mission data.

\section{Summary and Conclusions}  %% \conclusions[modified heading if necessary]
In this work, we have simulated the effects of mass redistribution within the atmosphere, terrestrial water storage, and glaciers over Europe and the resulting geopotential change on optical clock comparisons in a possible future clock network, which would be connected with colocated GNSS receivers. We went beyond previous investigations of these effects \citep{voigt_time-variable_2016}, and constructed three error scenarios for such a network, that also account for GNSS height control, in order to analyse which effects would be observable against the noise level. Hydrologic signals such as groundwater storage changes are comparably small over Europe and thus would hardly be seen under the most pessimistic scenario 1, i.e. when noise levels correspond to $10^{-18}$. At monthly and longer timescales however, they could be observed under scenarios 2 and 3, i.e. assuming uncertainties of $10^{-19}$ and $10^{-20}$, respectively, can be met for clocks and uncertainties of less than a millimeter for GNSS. As expected, we find that the influence of atmospheric mass variability is larger than for hydrological mass variability, with weekly and longer frequencies being detectable even under scenario 1, while in scenario 2 and 3 clock comparisons via a fibre network without loss of accuracy would enable one to observe even most of the daily signal. In contrast we find that glacier mass change in the European Alps and in the Scandinavian mountains would not contribute significantly to the clocks' observed signal. However, we note that in our scenarios clocks were assumed to be operated at existing national metrology laboratories and were not assumed to be transported and operated at designated observatories close to glaciers.
\par
The GRACE and GRACE-FO missions observe in principle the same potential variations, due mass redistribution within the same compartments of the Earth system as considered here; apart from the direct effect of vertical land motion on potential, which we here suggest would have to be corrected for, using vertical height change measurements from colocated GNSS. At this point we can conclude that a network of optical clocks could provide a new tool for GRACE validation. However, GNSS uncertainty would inevitably place limits on solving the geopotential change from time-differencing clock comparisons over long distances; clock comparisons with an uncertainty of $10^{-20}$ fractional frequency difference do not yield a significant gain compared to comparisons at $10^{-19}$, if a relative vertical motion of the clocks cannot be monitored beyond the 1 mm limit. Still, even if vertical height control is going to be a limiting factor for clock comparisons, the number of clocks can still be increased, which would benefit the network in particular in detecting variations of smaller spatial scale.
\par
We have not included non-tidal ocean loading in our simulations since most stations of our hypothetical network are located inland and for observing sea level changes and ocean loading, we argue that one would rather devise a network along coastal sites e.g. colocated with tide gauges. Furthermore, here we only looked at daily and longer variations. But, unless clock readings can be averaged over the whole day, tidal and nontidal subdaily variations would need to be considered by applying correction models due to aliasing effects.
\par
In this contribution we have assumed that time-differencing clock comparisons in a regional network would be analysed with respect to the GRACE data similar to how the superconducting gravimeter community works: Time-series of spectrally limited GRACE mass change are directly compared at instrument locations to measured time series which are to be corrected for local effects. We hypothesize that this correction is much easier to derive for clock measurements since they refer to the geoid while gravimeters refer to gravity, but one could nevertheless resort to common-mode isolation techniques such as Empirical Orthogonal Function analysis (EOF, \citealp{crossley_comparison_2012, van_camp_quest_2014}). Eventually, local effect such as groundwater table variations, snow depth or barometric pressure changes would have to be monitored.
\par
We conclude that in order to validate a future gravity mission that is about five times as precise as GRACE at monthly resolution, a clock network with a comparison error of maximally 2 mm geoid height would be needed, i.e. roughly our assumed scenario 2. It could be used for validation of the higher spatial scales of the gravity mission, i.e. 200 to 300 km.
For the monitoring of terrestrial water storage anomalies as an essential climate variable, proposed to be measured with $10$ to $20$ mm accuracy at 300 km spatial resolution, the clock network error level would need to be at the level of our assumed scenario 3, i.e. below $0.5$ mm geoid height.
At daily to weekly scales, the network of fibre-connected clocks that we consider here could additionally observe atmospheric mass changes under scenarios 2 and 3, and it could resolve hydrological mass changes in Europe under scenario 3, thus adding to the water cycle monitoring of a satellite gravity mission at daily temporal resolution.

\begin{acknowledgments}
The research leading to these results has received funding from the European Union’s Horizon 2020 research and innovation programme under Grant Agreement No. 951886 (CLONETS-DS).
We acknowledge partial support by the German Science Foundation DFG via the Research Unit NEROGRAV (KU1207/29-1).
Furthermore, we would like to thank Anne Springer and Anna Klos for helpful and interesting discussions.
\end{acknowledgments}

\dataavailability{All data used in this work can be received from the authors upon reasonable request.}

\bibliography{GJI}

\label{lastpage}

\end{document}